\documentclass[12pt]{article}
\usepackage{amssymb}
\usepackage{amsfonts}
\usepackage{amsmath}
\allowdisplaybreaks

\begin{document}

\author{C. Bizdadea\thanks{%
E-mail address: bizdadea@central.ucv.ro}, E. M. Cioroianu\thanks{%
E-mail address: manache@central.ucv.ro}, S. O. Saliu\thanks{%
E-mail address: osaliu@central.ucv.ro} \\
Faculty of Physics, University of Craiova\\
13 Al. I. Cuza Str., Craiova 200585, Romania\\
E. M. B\u{a}b\u{a}l\^{\i}c\thanks{%
E-mail address: mbabalic@central.ucv.ro}\\
Department of Theoretical Physics\\
Horia Hulubei National Institute of Physics and\\
Nuclear Engineering\\
PO Box MG-6, Bucharest, Magurele 077125, Romania}
\title{Yes-go cross-couplings in collections of tensor fields \\
with mixed symmetries of the type $(3,1)$ and $(2,2)$}
\date{}
\maketitle

\begin{abstract}
Under the hypotheses of analyticity, locality, Lorentz covariance, and
Poincar\'{e} invariance of the deformations, combined with the requirement
that the interaction vertices contain at most two space-time derivatives of
the fields, we investigate the consistent cross-couplings between two
collections of tensor fields with the mixed symmetries of the type $(3,1)$ and $(2,2)$.
The computations are done
with the help of the deformation theory based on a cohomological approach
in the context of the antifield-BRST formalism. Our results can be
synthesized in: 1. there appear consistent cross-couplings between the two
types of field collections at order one and two in the coupling constant
such that some of the gauge generators and of the reducibility functions are
deformed, and 2. the existence or not of cross-couplings among different
fields with the mixed symmetry of the Riemann tensor depends on the
indefinite or respectively positive-definite behaviour of the quadratic form
defined by the kinetic terms from the free Lagrangian.


\end{abstract}

\section{Introduction}

Tensor fields characterized by a mixed Young symmetry type (neither
completely antisymmetric nor fully symmetric)~\cite%
{curt,curt1,aul,labast,labast1,burd} attracted the attention lately on some
important issues, like the dual formulation of field theories of spin two or
higher~\cite{dualsp1,dualsp2,dualsp2a,dualsp2b,dualsp3,dualsp4,dualsp5}, a
Lagrangian first-order approach~\cite{zinov2} to some classes of massless
mixed symmetry-type tensor gauge fields, suggestively resembling to the
tetrad formalism of General Relativity, or the derivation of some exotic
gravitational interactions~\cite{boulangerCQG,ancoPRD}.

There exist in fact three different dual formulations of linearized gravity
in $D$ dimensions: the Pauli--Fierz description~\cite{pf1,pf2}, the version
based on a massless tensor field with the mixed symmetry $(D-3,1)$~\cite%
{aul,dualsp2,7}, and the formulation in terms of a massless tensor field
with the mixed symmetry $(D-3,D-3)$~\cite{9,kk}. The last two versions are
obtained by dualizing on one and respectively on both indices the
Pauli--Fierz field~\cite{dualsp1}. These dual formulations in terms of mixed
symmetry tensor gauge fields have been systematically investigated from the
perspective of $M$-theory~\cite{mth1,mth2,mth3}.

An important matter related to the dual formulations of linearized gravity
is the study of their consistent interactions, among themselves as well as
with other gauge theories. The most efficient approach to this problem is
the cohomological one, based on the deformation of the solution to the
master equation~\cite{def}. Since the mixed symmetry tensor fields involved
in dual formulations of linearized gravity allow no self-interactions, it
was believed that they are also rigid under the introduction of couplings to
other gauge theories. Nevertheless, recent results prove the contrary. For
instance, it was shown that some theories with massless tensor fields
exhibiting the mixed symmetry $(k,1)$ can be consistently coupled to a
vector field ($k=3$)~\cite{prd06}, to an arbitrary $p$-form ($k=3$)~\cite%
{jpa08}, to a topological BF model ($k=2$)~\cite{epjc09}, and to a massless
tensor field with the mixed symmetry of the Riemann tensor ($k=3$)~\cite%
{fort09}. There is a revived interest in the construction of dual gravity
theories, which led to several new results, viz. a dual formulation of
linearized gravity in first order tetrad formalism in arbitrary dimensions
within the path integral framework~\cite{sivakumar} or a reformulation of
non-linear Einstein gravity in terms of the dual graviton together with the
ordinary metric and a shift gauge field~\cite{boulangerhohm}.

A major result concerning spin-two fields within the standard formulation of
Einstein--Hilbert gravity is the impossibility of cross-couplings in
multi-graviton theories, either direct~\cite{multi} or intermediated by a
scalar field~\cite{multi}, a Dirac spinor~\cite{noijhepdirac}, a massive
Rarita--Schwinger field~\cite{noiRS}, or a massless $p$-form~\cite{noiNPB}.
The same no-go outcome has occurred at the level of multi-Weyl graviton
theories~\cite{marcweyl,noiweyl} and also in relation with dual formulations
of linearized gravity~\cite{lingr,jpa08}. These no-go results on
multi-graviton theories are important since they provide new arguments for
ruling out $N>8$ extended supergravity theories, as they would involve more
than one graviton.

The aim of this paper is to combine the study of consistent interactions
between two different dual formulations of linearized gravity with the
analysis of cross-couplings in collections of such dual multi-graviton
theories. More precisely, we generate all consistent interactions in a
collection of massless tensor fields with the mixed symmetry $(3,1)$, $%
\left\{ t_{\lambda \mu \nu |\kappa }^{A}\right\} _{A=\overline{1,N}}$, and a
collection of massless tensor fields with the mixed symmetry of the Riemann
tensor, $\left\{ r_{\mu \nu |\kappa \beta }^{a}\right\} _{a=\overline{1,n}}$%
. Special attention will be paid to the existence of cross-couplings among
different spin-two fields (with the mixed symmetry of the Riemann tensor)
intermediated by the presence of tensor fields with the mixed symmetry $%
(3,1) $. Our analysis relies on the deformation of the solution to the
master equation by means of cohomological techniques with the help of the
local BRST cohomology, whose component in a single $(3,1)$ sector has been
reported in detail in~\cite{t31jhep} and in a single $(2,2)$ sector has been
investigated in~\cite{r22,r22th}. The self-interactions in a collection of
tensor fields with the mixed symmetry $(3,1)$ and respectively $(2,2)$ has
been approached in~\cite{PAUC09}. We work in the standard hypotheses on the
deformations: analyticity in the coupling constant, locality, Lorentz
covariance, Poincar\'{e} invariance, and preservation of the number of
derivatives on each field (derivative order assumption). The derivative
order assumption is translated here into the requirement that the
interaction vertices contain at most two space-time derivatives acting on
the fields at all orders in the coupling constant.

We show that there exists a case where the deformed solution to the master
equation outputs non-trivial cross-couplings. It stops at order two in the
coupling constant and is defined on a space-time of dimension $D=6$, i.e.
precisely the dimension where the free fields with the mixed symmetry $(3,1)$
become dual to the linearized limit of Hilbert--Einstein gravity. The
interacting Lagrangian action contains only mixing-component terms of order
one and two in the coupling constant. Both the gauge transformations and
first-order reducibility functions of the tensor fields $(3,1)$ are modified
at order one in the coupling constant with terms characteristic to the $%
(2,2) $ sector. On the contrary, the tensor fields with the mixed symmetry $%
(2,2)$ remain rigid at the level of both gauge transformations and
reducibility functions. The gauge algebra and the reducibility structure of
order two are not modified during the deformation procedure, being the same
like in the case of the starting free action. The most important result is
that the existence of cross-couplings among different fields with the mixed
symmetry of the Riemann tensor is essentially dictated by the behaviour of
the metric tensor in the inner space of collection indices $a=\overline{1,n}$%
, $\hat{k}=\left( k_{ab}\right) $ (the quadratic form defined by the kinetic
terms from the free Lagrangian density for the fields $\left\{ r_{\mu \nu
|\kappa \beta }^{a}\right\} _{a=\overline{1,n}}$). Thus, if $\hat{k}$ is
positive-definite, then there appear no cross-couplings among different
fields from the collection $\left\{ r_{\mu \nu |\kappa \beta }^{a}\right\}
_{a=\overline{1,n}}$. On the contrary, if $\hat{k}$ is indefinite, then
there are allowed cross-couplings among different fields from this
collection.

\section{Brief review of the deformation procedure\label{briefrev}}

There are three main types of consistent interactions that can be added to a
given gauge theory: (i) the first type deforms only the Lagrangian action,
but not its gauge transformations, (ii) the second kind modifies both the
action and its transformations, but not the gauge algebra, and (iii) the
third, and certainly most interesting category, changes everything, namely,
the action, its gauge symmetries, and the accompanying algebra.

The reformulation of the problem of consistent deformations of a given
action and of its gauge symmetries in the antifield-BRST setting is based on
the observation that if a deformation of the classical theory can be
consistently constructed, then the solution $S$ to the master equation for
the initial theory can be deformed into the solution $\bar{S}$ of the master
equation for the interacting theory
\begin{eqnarray}
S &\longrightarrow &\bar{S}=S+gS_{1}+g^{2}S_{2}+g^{3}S_{3}+g^{4}S_{4}+\cdots
,  \label{ec39} \\
\left( S,S\right) =0 &\longrightarrow &\left( \bar{S},\bar{S}\right) =0.
\label{ec39b}
\end{eqnarray}%
The projection of (\ref{ec39b}) for $\bar{S}$ on the various powers of the
coupling constant induces the following tower of equations:
\begin{eqnarray}
g^{0} &:&\left( S,S\right) =0,  \label{ec40} \\
g^{1} &:&\left( S_{1},S\right) =0,  \label{ec41} \\
g^{2} &:&\left( S_{2},S\right) +\frac{1}{2}\left( S_{1},S_{1}\right) =0,
\label{ec42} \\
g^{3} &:&\left( S_{3},S\right) +\left( S_{1},S_{2}\right) =0,  \label{ec43}
\\
g^{4} &:&\left( S_{4},S\right) +\left( S_{1},S_{3}\right) +\frac{1}{2}\left(
S_{2},S_{2}\right) =0,  \label{ec44} \\
&&\vdots  \notag
\end{eqnarray}%
The first equation is satisfied by hypothesis. The second one governs the
first-order deformation of the solution to the master equation, $S_{1}$, and
it expresses the fact that $S_{1}$ is a BRST co-cycle, $sS_{1}=0$, and hence
it exists and is local. The remaining equations are responsible for the
higher-order deformations of the solution to the master equation. No
obstructions arise in finding solutions to them as long as no further
restrictions, such as space-time locality, are imposed. Obviously, only
non-trivial first-order deformations should be considered, since trivial
ones ($S_{1}=sB$) lead to trivial deformations of the initial theory, and
can be eliminated by convenient redefinitions of the fields. Ignoring the
trivial deformations, it follows that $S_{1}$ is a non-trivial
BRST-observable, $S_{1}\in H^{0}\left( s\right) $ (where $H^{0}\left(
s\right) $ denotes the cohomology space of the BRST differential in ghost
number zero). Once the deformation equations ((\ref{ec41})--(\ref{ec44}),
etc.) have been solved by means of specific cohomological techniques, from
the consistent, non-trivial deformed solution to the master equation one can
extract all the information on the gauge structure of the resulting
interacting theory.

\section{Free model: Lagrangian formulation and BRST symmetry \label%
{colfreer22t31}}

We start from a free theory in $D\geq 5$ that describes two finite
collections of massless tensor fields with the mixed symmetries $(3,1)$ and
respectively $(2,2)$
\begin{equation}
S_{0}\left[ t_{\lambda \mu \nu |\kappa }^{A},r_{\mu \nu |\kappa \beta }^{a}%
\right] =S_{0}^{\mathrm{t}}\left[ t_{\lambda \mu \nu |\kappa }^{A}\right]
+S_{0}^{\mathrm{r}}\left[ r_{\mu \nu |\kappa \beta }^{a}\right] ,
\label{colrt1b}
\end{equation}%
where
\begin{eqnarray}
S_{0}^{\mathrm{t}}\left[ t_{\lambda \mu \nu |\kappa }^{A}\right] &=&\int
\left\{ \frac{1}{2}\left[ \left( \partial ^{\rho }t_{A}^{\lambda \mu \nu
|\kappa }\right) \left( \partial _{\rho }t_{\lambda \mu \nu |\kappa
}^{A}\right) -\left( \partial _{\kappa }t_{A}^{\lambda \mu \nu |\kappa
}\right) \left( \partial ^{\beta }t_{\lambda \mu \nu |\beta }^{A}\right) %
\right] \right.  \notag \\
&&-\frac{3}{2}\left[ \left( \partial _{\lambda }t_{A}^{\lambda \mu \nu
|\kappa }\right) \left( \partial ^{\rho }t_{\rho \mu \nu |\kappa
}^{A}\right) +\left( \partial ^{\rho }t_{A}^{\lambda \mu }\right) \left(
\partial _{\rho }t_{\lambda \mu }^{A}\right) \right]  \notag \\
&&\left. +3\left[ \left( \partial _{\kappa }t_{A}^{\lambda \mu \nu |\kappa
}\right) \left( \partial _{\lambda }t_{\mu \nu }^{A}\right) +\left( \partial
_{\rho }t_{A}^{\rho \mu }\right) \left( \partial ^{\lambda }t_{\lambda \mu
}^{A}\right) \right] \right\} d^{D}x,  \label{colt1}
\end{eqnarray}%
\begin{eqnarray}
S_{0}^{\mathrm{r}}\left[ r_{\mu \nu |\kappa \beta }^{a}\right] &=&\int
\left\{ -\frac{1}{2}\left[ \left( \partial _{\mu }r_{a}^{\mu \nu |\kappa
\beta }\right) \left( \partial ^{\lambda }r_{\lambda \nu |\kappa \beta
}^{a}\right) +\left( \partial ^{\lambda }r_{a}^{\nu \beta }\right) \left(
\partial _{\lambda }r_{\nu \beta }^{a}\right) \right. \right.  \notag \\
&&\left. +\left( \partial _{\nu }r_{a}^{\nu \beta }\right) \left( \partial
_{\beta }r^{a}\right) \right] +\frac{1}{8}\left[ \left( \partial ^{\lambda
}r_{a}^{\mu \nu |\kappa \beta }\right) \left( \partial _{\lambda }r_{\mu \nu
|\kappa \beta }^{a}\right) +\left( \partial ^{\lambda }r_{a}\right) \left(
\partial _{\lambda }r^{a}\right) \right]  \notag \\
&&\left. -\left( \partial _{\mu }r_{a}^{\mu \nu |\kappa \beta }\right)
\left( \partial _{\beta }r_{\nu \kappa }^{a}\right) +\left( \partial _{\nu
}r_{a}^{\nu \beta }\right) \left( \partial ^{\lambda }r_{\lambda \beta
}^{a}\right) \right\} d^{D}x.  \label{colr1}
\end{eqnarray}%
Everywhere in this paper we employ the flat Minkowski metric of `mostly
plus' signature $\sigma ^{\mu \nu }=\sigma _{\mu \nu }=(-++++\ldots )$. The
uppercase indices $A$, $B$, etc. stand for the collection indices of the
fields with the mixed symmetry $(3,1)$ and are assumed to take discrete
values: $1$, $2$, $\ldots $, $N$. They are lowered with a symmetric,
constant, and invertible matrix, of elements $k_{AB}$, and are raised with
the help of the elements $k^{AB}$ of its inverse. This means that $%
t_{A}^{\lambda \mu \nu |\kappa }=k_{AB}t^{B\lambda \mu \nu |\kappa }$ and $%
t_{\lambda \mu \nu |\kappa }^{A}=k^{AB}t_{B\lambda \mu \nu |\kappa }$. Each
field $t_{\lambda \mu \nu |\kappa }^{A}$ is completely antisymmetric in its
first three (Lorentz) indices and satisfies the identity $t_{\left[ \lambda
\mu \nu |\kappa \right] }^{A}\equiv 0$. Here and in the sequel the notation $%
[\lambda \ldots \kappa ]$ signifies complete antisymmetry with respect to
the (Lorentz) indices between brackets, with the conventions that the
minimum number of terms is always used and the result is never divided by
the number of terms. The notation $t_{\lambda \mu }^{A}$ signifies the trace
of $t_{\lambda \mu \nu |\kappa }^{A}$, defined by $t_{\lambda \mu
}^{A}=\sigma ^{\nu \kappa }t_{\lambda \mu \nu |\kappa }^{A}$. The trace
components define an antisymmetric tensor, $t_{\lambda \mu }^{A}=-t_{\mu
\lambda }^{A}$. The lowercase indices $a$, $b$, etc. stand for the
collection indices of the fields with the mixed symmetry $(2,2)$ and are
assumed to take the discrete values $1$, $2$, $\ldots $, $n$. They are
lowered with a symmetric, constant, and invertible matrix, of elements $%
k_{ab}$, and are raised with the help of the elements $k^{ab}$ of its
inverse, such that $r_{a}^{\mu \nu |\kappa \beta }=k_{ab}r^{b\mu \nu |\kappa
\beta }$ and $r_{\lambda \nu |\kappa \beta }^{a}=k^{ab}r_{b\lambda \nu
|\kappa \beta }$. Each tensor field $r_{\mu \nu |\kappa \beta }^{a}$ is
separately antisymmetric in the pairs $\left\{ \mu ,\nu \right\} $ and $%
\left\{ \kappa ,\beta \right\} $, is symmetric under their permutation ($%
\left\{ \mu ,\nu \right\} \longleftrightarrow \left\{ \kappa ,\beta \right\}
$), and satisfies the identity $r_{\left[ \mu \nu |\kappa \right] \beta
}^{a}\equiv 0$. The notations $r_{\nu \beta }^{a}$ signify the traces of $%
r_{\mu \nu |\kappa \beta }^{a}$, $r_{\nu \beta }^{a}=\sigma ^{\mu \kappa
}r_{\mu \nu |\kappa \beta }^{a}$, which are symmetric, $r_{\nu \beta
}^{a}=r_{\beta \nu }^{a}$, while $r^{a}$ represent their double traces, $%
r^{a}=\sigma ^{\nu \beta }r_{\nu \beta }^{a}$, which are scalars.

A generating set of gauge transformations of action (\ref{colrt1b}) can be
taken as
\begin{eqnarray}
\delta _{\epsilon ,\chi }t_{\lambda \mu \nu |\kappa }^{A} &=&3\epsilon
_{\lambda \mu \nu ,\kappa }^{A}+\partial _{\left[ \lambda \right. }\epsilon
_{\left. \mu \nu \right] \kappa }^{A}+\partial _{\left[ \lambda \right.
}\chi _{\left. \mu \nu \right] |\kappa }^{A},  \label{colt7} \\
\delta _{\xi }r_{\mu \nu |\kappa \beta }^{a} &=&\xi _{\kappa \beta |\left[
\nu ,\mu \right] }^{a}+\xi _{\mu \nu |\left[ \beta ,\kappa \right] }^{a},
\label{colr8}
\end{eqnarray}%
where we used the standard notation $f_{,\mu }=\partial f/\partial x^{\mu }$%
. All the gauge parameters are bosonic, with $\epsilon _{\lambda \mu \nu
}^{A}$ completely antisymmetric and $\chi _{\mu \nu |\kappa }^{A}$ together
with $\xi _{\mu \nu |\kappa }^{a}$ defining two collections of tensor fields
with the mixed symmetry $(2,1)$. The former gauge transformations, (\ref%
{colt7}), are off-shell, second-order reducible in the space of all field
histories, the associated gauge algebra being Abelian (see~\cite%
{t31jhep,PAUC09}), while the gauge symmetries (\ref{colr8}) are off-shell,
first-order reducible, the corresponding algebra being also Abelian (see~%
\cite{r22,PAUC09}). It follows that the free theory (\ref{colrt1b}) is a
linear gauge theory with the Cauchy order equal to four. The simplest gauge
invariant quantities are precisely the curvature tensors%
\begin{equation}
K_{A}^{\lambda \mu \nu \xi |\kappa \beta }=t_{A}^{\left[ \mu \nu \xi
,\lambda \right] |\left[ \beta ,\kappa \right] },\quad F_{\mu \nu \lambda
|\kappa \beta \gamma }^{a}=r_{\left[ \mu \nu ,\lambda \right] |\left[ \kappa
\beta ,\gamma \right] }^{a},  \label{curvatures}
\end{equation}%
and their space-time derivatives. It is easy to check that they display the
mixed symmetry $(4,2)$ and $(3,3)$ respectively.

The construction of the BRST symmetry for the free model under study debuts
with the identification of the algebra on which the BRST differential $s$
acts. The ghost spectrum comprises the fermionic ghosts $\left\{ \eta
_{\lambda \mu \nu }^{A},\mathcal{G}_{\mu \nu |\kappa }^{A},\mathcal{C}_{\mu
\nu |\kappa }^{a}\right\} $ respectively associated with the gauge
parameters $\left\{ \epsilon _{\lambda \mu \nu }^{A},\chi _{\mu \nu |\kappa
}^{A},\xi _{\mu \nu |\kappa }^{a}\right\} $ from (\ref{colt7}) and (\ref%
{colr8}), the bosonic ghosts for ghosts $\left\{ C_{\mu \nu }^{A},G_{\nu
\kappa }^{A},\mathcal{C}_{\mu \nu }^{a}\right\} $ due to the first-order
reducibility, and the fermionic ghosts for ghosts for ghosts $\left\{ C_{\nu
}^{A}\right\} $ corresponding to the maximum reducibility order (two). We
ask that $\eta _{\lambda \mu \nu }^{A}$, $C_{\mu \nu }^{A}$, and $\mathcal{C}%
_{\mu \nu }^{a}$ are completely antisymmetric, $\mathcal{G}_{\mu \nu |\kappa
}^{A}$ and $\mathcal{C}_{\mu \nu |\kappa }^{a}$ exhibit the mixed symmetry $%
(2,1)$, and $G_{\nu \kappa }^{A}$ are symmetric. The antifield spectrum
comprises the antifields $\left\{ t_{A}^{\ast \lambda \mu \nu |\kappa
},r_{a}^{\ast \mu \nu |\kappa \beta }\right\} $ associated with the original
fields and those corresponding to the ghosts, $\left\{ \eta _{A}^{\ast
\lambda \mu \nu },\mathcal{G}_{A}^{\ast \mu \nu |\kappa },\mathcal{C}%
_{a}^{\ast \mu \nu |\kappa }\right\} $, $\left\{ C_{A}^{\ast \mu \nu
},G_{A}^{\ast \nu \kappa },\mathcal{C}_{a}^{\ast \mu \nu }\right\} $, and $%
\left\{ C_{A}^{\ast \nu }\right\} $. The antifields are required to satisfy
the same symmetry, antisymmetry, or mixed symmetry properties like the
corresponding fields/ghosts. Related to the traces of the antifields, we
will use the notations $t_{A}^{\ast \lambda \mu }=\sigma _{\nu \kappa
}t_{A}^{\ast \lambda \mu \nu |\kappa }$, $r_{a}^{\ast \nu \beta }=\sigma
_{\mu \kappa }r_{a}^{\ast \mu \nu |\kappa \beta }$, and $r_{a}^{\ast
}=\sigma _{\nu \beta }r_{a}^{\ast \nu \beta }$.

Since both the gauge generators and reducibility functions for this model
are field-independent, it follows that the BRST differential $s$ simply
reduces to
\begin{equation}
s=\delta +\gamma ,  \label{colBRST}
\end{equation}%
where $\delta $ represents the Koszul--Tate differential, graded by the
antighost number $\mathrm{agh}$ ($\mathrm{agh}\left( \delta \right) =-1$),
and $\gamma $ stands for the exterior longitudinal differential, whose
degree is named pure ghost number $\mathrm{pgh}$ ($\mathrm{pgh}\left( \gamma
\right) =1$). These two degrees do not interfere ($\mathrm{agh}\left( \gamma
\right) =0$, $\mathrm{pgh}\left( \delta \right) =0$). The overall degree
that grades the BRST complex is known as the ghost number ($\mathrm{gh}$)
and is defined like the difference between the pure ghost number and the
antighost number, such that $\mathrm{gh}\left( s\right) =\mathrm{gh}\left(
\delta \right) =\mathrm{gh}\left( \gamma \right) =1$. According to the
standard rules of the BRST method, the corresponding degrees of the
generators from the BRST complex are valued like%
\begin{gather*}
\mathrm{pgh}\left( t_{\lambda \mu \nu |\kappa }^{A}\right) =0=\mathrm{pgh}%
\left( r_{\mu \nu |\kappa \beta }^{a}\right) , \\
\mathrm{pgh}\left( \eta _{\lambda \mu \nu }^{A}\right) =\mathrm{pgh}\left(
\mathcal{G}_{\mu \nu |\kappa }^{A}\right) =\mathrm{pgh}\left( \mathcal{C}%
_{\mu \nu |\kappa }^{a}\right) =1, \\
\mathrm{pgh}\left( C_{\mu \nu }^{A}\right) =\mathrm{pgh}\left( G_{\nu \kappa
}^{A}\right) =\mathrm{pgh}\left( \mathcal{C}_{\mu \nu }^{a}\right) =2,\quad
\mathrm{pgh}\left( C_{\nu }^{A}\right) =3, \\
\mathrm{pgh}\left( \Phi _{\Delta }^{\ast }\right) =0=\mathrm{agh}\left( \Phi
^{\Delta }\right) , \\
\mathrm{agh}\left( t_{A}^{\ast \lambda \mu \nu |\kappa }\right) =1=\mathrm{%
agh}\left( r_{a}^{\ast \mu \nu |\kappa \beta }\right) , \\
\mathrm{agh}\left( \eta _{A}^{\ast \lambda \mu \nu }\right) =\mathrm{agh}%
\left( \mathcal{G}_{A}^{\ast \mu \nu |\kappa }\right) =\mathrm{agh}\left(
\mathcal{C}_{a}^{\ast \mu \nu |\kappa }\right) =2, \\
\mathrm{agh}\left( C_{A}^{\ast \mu \nu }\right) =\mathrm{agh}\left(
G_{A}^{\ast \nu \kappa }\right) =\mathrm{agh}\left( \mathcal{C}_{a}^{\ast
\mu \nu }\right) =3,\quad \mathrm{agh}\left( C_{A}^{\ast \nu }\right) =4,
\end{gather*}%
where we made the notations%
\begin{eqnarray}
\Phi ^{\Delta } &=&\left\{ t_{\lambda \mu \nu |\kappa }^{A},r_{\mu \nu
|\kappa \beta }^{a},\eta _{\lambda \mu \nu }^{A},\mathcal{G}_{\mu \nu
|\kappa }^{A},\mathcal{C}_{\mu \nu |\kappa }^{a},C_{\mu \nu }^{A},G_{\nu
\kappa }^{A},\mathcal{C}_{\mu \nu }^{a},C_{\nu }^{A}\right\} ,
\label{fields} \\
\Phi _{\Delta }^{\ast } &=&\left\{ t_{A}^{\ast \lambda \mu \nu |\kappa
},r_{a}^{\ast \mu \nu |\kappa \beta },\eta _{A}^{\ast \lambda \mu \nu },%
\mathcal{G}_{A}^{\ast \mu \nu |\kappa },\mathcal{C}_{a}^{\ast \mu \nu
|\kappa },C_{A}^{\ast \mu \nu },G_{A}^{\ast \nu \kappa },\mathcal{C}%
_{a}^{\ast \mu \nu },C_{A}^{\ast \nu }\right\} .  \label{antif}
\end{eqnarray}%
The Koszul--Tate differential is imposed to realize a homological resolution
of the algebra of smooth functions defined on the stationary surface of
field equations, while the exterior longitudinal differential is related to
the gauge symmetries (see relations (\ref{colt7}) and (\ref{colr8})) of
action (\ref{colrt1b}) through its cohomology at pure ghost number zero
computed in the cohomology of $\delta $, which is required to be the algebra
of physical observables for the free model under consideration. The actions
of $\delta $ and $\gamma $ on the generators from the BRST complex, which
enforce all the above mentioned properties, are given by%
\begin{gather}
\gamma t_{\lambda \mu \nu |\kappa }^{A}=-3\partial _{\left[ \lambda \right.
}\eta _{\left. \mu \nu \kappa \right] }^{A}+4\partial _{\left[ \lambda
\right. }\eta _{\left. \mu \nu \right] \kappa }^{A}+\partial _{\left[
\lambda \right. }\mathcal{G}_{\left. \mu \nu \right] |\kappa }^{A},
\label{colt49} \\
\gamma r_{\mu \nu |\kappa \beta }^{a}=\partial _{\mu }\mathcal{C}_{\kappa
\beta |\nu }^{a}-\partial _{\nu }\mathcal{C}_{\kappa \beta |\mu
}^{a}+\partial _{\kappa }\mathcal{C}_{\mu \nu |\beta }^{a}-\partial _{\beta }%
\mathcal{C}_{\mu \nu |\kappa }^{a},  \label{colr50} \\
\gamma \eta _{\lambda \mu \nu }^{A}=-\frac{1}{2}\partial _{\left[ \lambda
\right. }C_{\left. \mu \nu \right] }^{A},  \label{colt50} \\
\gamma \mathcal{G}_{\mu \nu |\kappa }^{A}=2\partial _{\left[ \mu \right.
}C_{\left. \nu \kappa \right] }^{A}-3\partial _{\left[ \mu \right.
}C_{\left. \nu \right] \kappa }^{A}+\partial _{\left[ \mu \right. }G_{\left.
\nu \right] \kappa }^{A},  \label{colt51} \\
\gamma \mathcal{C}_{\mu \nu |\kappa }^{a}=2\partial _{\kappa }\mathcal{C}%
_{\mu \nu }^{a}-\partial _{\left[ \mu \right. }\mathcal{C}_{\left. \nu %
\right] \kappa }^{a},\quad \gamma \mathcal{C}_{\mu \nu }^{a}=0,
\label{colr51} \\
\gamma C_{\mu \nu }^{A}=\partial _{\left[ \mu \right. }C_{\left. \nu \right]
}^{A},\quad \gamma G_{\nu \kappa }^{A}=-3\partial _{\left( \nu \right.
}C_{\left. \kappa \right) }^{A},\quad \gamma C_{\nu }^{A}=0,  \label{colt52}
\\
\gamma \Phi _{\Delta }^{\ast }=0=\delta \Phi ^{\Delta },  \label{colrcolt} \\
\delta t_{A}^{\ast \lambda \mu \nu |\kappa }=T_{A}^{\lambda \mu \nu |\kappa
},\quad \delta \eta _{A}^{\ast \lambda \mu \nu }=-4\partial _{\kappa
}t_{A}^{\ast \lambda \mu \nu |\kappa },  \label{colt55} \\
\delta \mathcal{G}_{A}^{\ast \mu \nu |\kappa }=-\partial _{\lambda }\left(
3t_{A}^{\ast \lambda \mu \nu |\kappa }-t_{A}^{\ast \mu \nu \kappa |\lambda
}\right) ,  \label{colt56} \\
\delta C_{A}^{\ast \mu \nu }=3\partial _{\lambda }\left( \mathcal{G}%
_{A}^{\ast \mu \nu |\lambda }-\frac{1}{2}\eta _{A}^{\ast \lambda \mu \nu
}\right) ,\quad \delta G_{A}^{\ast \nu \kappa }=\partial _{\mu }\mathcal{G}%
_{A}^{\ast \mu \left( \nu |\kappa \right) },  \label{colt57} \\
\delta C_{A}^{\ast \nu }=6\partial _{\mu }\left( G_{A}^{\ast \mu \nu }-\frac{%
1}{3}C_{A}^{\ast \mu \nu }\right) ,  \label{colt58} \\
\delta r_{a}^{\ast \mu \nu |\kappa \beta }=\frac{1}{4}R_{a}^{\mu \nu |\kappa
\beta },\quad \delta \mathcal{C}_{a}^{\ast \kappa \beta |\nu }=-4\partial
_{\mu }r_{a}^{\ast \mu \nu |\kappa \beta },\quad \delta \mathcal{C}%
_{a}^{\ast \mu \nu }=3\partial _{\kappa }\mathcal{C}_{a}^{\ast \mu \nu
|\kappa },  \label{colr53}
\end{gather}%
where $T_{A}^{\lambda \mu \nu |\kappa }=-\delta S_{0}^{\mathrm{t}}/\delta
t_{\lambda \mu \nu |\kappa }^{A}$ and $\delta S_{0}^{\mathrm{r}}/\delta
r_{a}^{\mu \nu |\kappa \beta }\equiv -\left( 1/4\right) R_{\mu \nu |\kappa
\beta }^{a}$. By convention, we take $\delta $ and $\gamma $ to act like
right derivations.

We note that the action of the Koszul--Tate differential on the antifields
with the antighost number equal to two and respectively three from the $%
(3,1) $ sector gains a simpler expression if we perform the changes of
variables%
\begin{equation}
\mathcal{G}_{A}^{\prime \ast \mu \nu |\kappa }=\mathcal{G}_{A}^{\ast \mu \nu
|\kappa }+\frac{1}{4}\eta _{A}^{\ast \mu \nu \kappa },\quad G_{A}^{\prime
\ast \nu \kappa }=G_{A}^{\ast \nu \kappa }-\frac{1}{3}C_{A}^{\ast \nu \kappa
}.  \label{colt58a}
\end{equation}%
The antifields $\mathcal{G}_{A}^{\prime \ast \mu \nu |\kappa }$ are still
antisymmetric in their first two indices, but do not fulfill the identity $%
\mathcal{G}_{A}^{\prime \ast \left[ \mu \nu |\kappa \right] }\equiv 0$, and $%
G_{A}^{\prime \ast \nu \kappa }$ have no definite symmetry or antisymmetry
properties. With the help of relations (\ref{colt55})--(\ref{colt58}), we
find that $\delta $ acts on the transformed antifields through the relations%
\begin{equation}
\delta \mathcal{G}_{A}^{\prime \ast \mu \nu |\kappa }=-3\partial _{\lambda
}t_{A}^{\ast \lambda \mu \nu |\kappa },\quad \delta G_{A}^{\prime \ast \nu
\kappa }=2\partial _{\mu }\mathcal{G}_{A}^{\prime \ast \mu \nu |\kappa
},\quad \delta C_{A}^{\ast \nu }=6\partial _{\mu }G_{A}^{\prime \ast \mu \nu
}.  \label{colt58b}
\end{equation}%
The same observation is valid with respect to $\gamma $ if we make the
changes
\begin{equation}
\mathcal{G}_{\mu \nu |\kappa }^{\prime A}=\mathcal{G}_{\mu \nu |\kappa
}^{A}+4\eta _{\mu \nu \kappa }^{A},\quad G_{\nu \kappa }^{\prime A}=G_{\nu
\kappa }^{A}-3C_{\nu \kappa }^{A},  \label{colt58ba}
\end{equation}%
in terms of which we can write
\begin{equation}
\gamma t_{\lambda \mu \nu |\kappa }^{A}=-\frac{1}{4}\partial _{\left[
\lambda \right. }\mathcal{G}_{\left. \mu \nu |\kappa \right] }^{\prime
A}+\partial _{\left[ \lambda \right. }\mathcal{G}_{\left. \mu \nu \right]
|\kappa }^{\prime A},\quad \gamma \mathcal{G}_{\mu \nu |\kappa }^{\prime
A}=\partial _{\left[ \mu \right. }G_{\left. \nu \right] \kappa }^{\prime
A},\quad \gamma G_{\nu \kappa }^{\prime A}=-6\partial _{\nu }C_{\kappa }^{A}.
\label{colt58bd}
\end{equation}%
Again, $\mathcal{G}_{\mu \nu |\kappa }^{\prime A}$ are antisymmetric in
their first two indices, but do not satisfy the identity $\mathcal{G}_{\left[
\mu \nu |\kappa \right] }^{\prime A}\equiv 0$, while $G_{\nu \kappa
}^{\prime A}$ have no definite symmetry or antisymmetry. We have
deliberately chosen the same notations for the transformed variables (\ref%
{colt58a}) and (\ref{colt58ba}) since they actually form pairs that are
conjugated in the antibracket%
\begin{equation*}
\left( \mathcal{G}_{\mu \nu |\kappa }^{\prime A},\mathcal{G}_{B}^{\prime
\ast \mu _{1}\nu _{1}|\kappa _{1}}\right) =\frac{1}{2}\delta _{B}^{A}\delta
_{\mu }^{\left[ \mu _{1}\right. }\delta _{\nu }^{\left. \nu _{1}\right]
}\delta _{\kappa }^{\kappa _{1}},\quad \left( G_{\nu \kappa }^{\prime
A},G_{B}^{\prime \ast \nu _{1}\kappa _{1}}\right) =\delta _{B}^{A}\delta
_{\nu }^{\nu _{1}}\delta _{\kappa }^{\kappa _{1}}.
\end{equation*}

The Lagrangian BRST differential admits a canonical action in a structure
named antibracket and defined by decreeing the fields/ghosts conjugated with
the corresponding antifields, $s\cdot =\left( \cdot ,S\right) $, where $%
\left( ,\right) $ signifies the antibracket and $S$ denotes the canonical
generator of the BRST symmetry. It is a bosonic functional of ghost number
zero, involving both field/ghost and antifield spectra, that obeys the
master equation $\left( S,S\right) =0$. The master equation is equivalent
with the second-order nilpotency of $s$, where its solution $S$ encodes the
entire gauge structure of the associated theory. Taking into account
formulas (\ref{colt49})--(\ref{colr53}) as well as the standard actions of $%
\delta $ and $\gamma $ in canonical form, we find that the complete solution
to the master equation for the free model under study is given by%
\begin{equation}
S=S^{\mathrm{t}}+S^{\mathrm{r}},  \label{colrt2}
\end{equation}%
where%
\begin{eqnarray}
S^{\mathrm{t}} &=&S_{0}^{\mathrm{t}}\left[ t_{\lambda \mu \nu |\kappa }^{A}%
\right] +\int \left[ t_{A}^{\ast \lambda \mu \nu |\kappa }\left( 3\partial
_{\kappa }\eta _{\lambda \mu \nu }^{A}+\partial _{\left[ \lambda \right.
}\eta _{\left. \mu \nu \right] \kappa }^{A}+\partial _{\left[ \lambda
\right. }\mathcal{G}_{\left. \mu \nu \right] |\kappa }^{A}\right) \right.
\notag \\
&&-\frac{1}{2}\eta _{A}^{\ast \lambda \mu \nu }\partial _{\left[ \lambda
\right. }C_{\left. \mu \nu \right] }^{A}+\mathcal{G}_{A}^{\ast \mu \nu
|\kappa }\left( 2\partial _{\kappa }C_{\mu \nu }^{A}-\partial _{\left[ \mu
\right. }C_{\left. \nu \right] \kappa }^{A}+\partial _{\left[ \mu \right.
}G_{\left. \nu \right] \kappa }^{A}\right)  \notag \\
&&\left. +C_{A}^{\ast \mu \nu }\partial _{\left[ \mu \right. }C_{\left. \nu %
\right] }^{A}-3G_{A}^{\ast \nu \kappa }\partial _{\left( \nu \right.
}C_{\left. \kappa \right) }^{A}\right] d^{D}x,  \label{colt60}
\end{eqnarray}
\begin{eqnarray}
S^{\mathrm{r}} &=&S_{0}^{\mathrm{r}}\left[ r_{\mu \nu |\kappa \beta }^{a}%
\right] +\int \left[ r_{a}^{\ast \mu \nu |\kappa \beta }\left( \partial
_{\mu }\mathcal{C}_{\kappa \beta |\nu }^{a}-\partial _{\nu }\mathcal{C}%
_{\kappa \beta |\mu }^{a}+\partial _{\kappa }\mathcal{C}_{\mu \nu |\beta
}^{a}-\partial _{\beta }\mathcal{C}_{\mu \nu |\kappa }^{a}\right) \right.
\notag \\
&&\left. +\mathcal{C}_{a}^{\ast \mu \nu |\kappa }\left( 2\partial _{\kappa }%
\mathcal{C}_{\mu \nu }^{a}-\partial _{\left[ \mu \right. }\mathcal{C}%
_{\left. \nu \right] \kappa }^{a}\right) \right] d^{D}x.  \label{colr55}
\end{eqnarray}

\section{Computation of basic cohomologies\label{colbasiccohr22t31}}

In the sequel we investigate the consistent couplings that can be added to
the free theory (\ref{colrt1b}) without modifying either the field spectrum
or the number of independent gauge invariances. In view of this we apply the
deformation procedure based on local BRST cohomology exposed in section \ref%
{briefrev} and solve equations (\ref{ec41})--(\ref{ec44}), etc. The
space-time locality of the deformations is ensured by working in the algebra
of local differential forms with coefficients that are polynomial functions
in the fields, ghosts, antifields, and their space-time derivatives (algebra
of local forms). In other words, the non-integrated density of the
first-order deformation, $a$, is assumed to be a polynomial function in all
these variables (algebra of local functions). The derivative order
assumption restricts the interaction Lagrangian to contain only interaction
vertices with maximum two space-time derivatives.

It is natural to decompose $a$ as a sum of three components%
\begin{equation}
a=a^{\mathrm{t}}+a^{\mathrm{r}}+a^{\mathrm{int}},  \label{coltv84}
\end{equation}%
where $a^{\mathrm{t}}$ denotes the part responsible for the
self-interactions of the fields $t_{\lambda \mu \nu |\kappa }^{A}$, $a^{%
\mathrm{r}}$ is related to the self-interactions of the fields $r_{\mu \nu
|\kappa \beta }^{a}$, and $a^{\mathrm{int}}$ signifies the component that
describes only the cross-couplings between $t_{\lambda \mu \nu |\kappa }^{A}$
and $r_{\mu \nu |\kappa \beta }^{a}$, so each term must mix the BRST
generators from the two sectors. According to decomposition (\ref{coltv84}),
equation $sa=\partial _{\mu }m^{\mu }$ becomes equivalent with three
equations%
\begin{equation}
sa^{\mathrm{t}}=\partial _{\mu }m_{\mathrm{t}}^{\mu },\quad sa^{\mathrm{r}%
}=\partial _{\mu }m_{\mathrm{r}}^{\mu },\quad sa^{\mathrm{int}}=\partial
_{\mu }m_{\mathrm{int}}^{\mu }.  \label{coltv85c}
\end{equation}%
The most general solutions to the first two equations from (\ref{coltv85c})
were approached in~\cite{PAUC09}, where it was shown that%
\begin{equation}
a^{\mathrm{t}}=0,\quad a^{\mathrm{r}}=c_{a}r^{a},  \label{colxz1}
\end{equation}%
with $c_{a}$ some arbitrary, real constants and $r^{a}$ the contractions of
order two of the fields $r_{\mu \nu |\kappa \beta }^{a}$. In the sequel we
approach the last equation from (\ref{coltv85c}).

Developing $a^{\mathrm{int}}$ according to the antighost number and assuming
that this expansion stops at a maximum, finite value $I$ of this degree, we
find that the equation $sa^{\mathrm{int}}=\partial _{\mu }m_{\mathrm{int}%
}^{\mu }$ becomes equivalent to the chain
\begin{eqnarray}
\gamma a_{I}^{\mathrm{int}} &=&\partial _{\mu }\overset{(I)}{m}_{\mathrm{int}%
}^{\mu },  \label{coltv65c} \\
\delta a_{I}^{\mathrm{int}}+\gamma a_{I-1}^{\mathrm{int}} &=&\partial _{\mu }%
\overset{(I-1)}{m}_{\mathrm{int}}^{\mu },  \label{coltv65d} \\
\delta a_{k}^{\mathrm{int}}+\gamma a_{k-1}^{\mathrm{int}} &=&\partial _{\mu }%
\overset{(k-1)}{m}_{\mathrm{int}}^{\mu },\quad I-1\geq k\geq 1.
\label{coltv65e}
\end{eqnarray}%
Equation (\ref{coltv65c}) can be replaced in strictly positive values of the
antighost number with%
\begin{equation}
\gamma a_{I}^{\mathrm{int}}=0,\quad \mathrm{agh}\left( a_{I}^{\mathrm{int}%
}\right) =I>0.  \label{coltv65f}
\end{equation}%
At this stage we notice that equation $sa^{\mathrm{int}}=\partial _{\mu }m_{%
\mathrm{int}}^{\mu }$ means that $a^{\mathrm{int}}d^{D}x\in H^{0,D}(s|d)$,
while equation (\ref{coltv65f}) shows that for $I>0$ $a_{I}^{\mathrm{int}%
}\in H^{\ast }\left( \gamma \right) $ (cohomology algebra of the exterior
longitudinal differential $\gamma $ computed in the algebra of local
functions mentioned in the above). Consequently, we need to compute $H^{\ast
}\left( \gamma \right) $. Combining the results inferred in~\cite{PAUC09}
on the cohomology algebra of the exterior longitudinal differential
in each sector, we obtain that the cohomology algebra $H^{\ast }(\gamma )$
computed in the algebra of local functions is generated on one hand by the
antifields (\ref{antif}), the curvature tensors (\ref{curvatures}), and
their space-time derivative and, on the other hand, by the ghosts or ghost
combinations $\mathcal{F}_{\lambda \mu \nu \kappa }^{A}$, $C_{\nu }^{A}$, $%
\mathcal{C}_{\mu \nu }^{a}$, and $\partial _{\left[ \mu \right. }\mathcal{C}%
_{\left. \nu \kappa \right] }^{a}$, where%
\begin{equation}
\mathcal{F}_{\lambda \mu \nu \kappa }^{A}=\partial _{\left[ \lambda \right.
}\eta _{\left. \mu \nu \kappa \right] }^{A}.  \label{defcalf}
\end{equation}%
Therefore, the general, local solution to equation (\ref{coltv65f}) is
expressed (up to trivial, $\gamma $-exact contributions) by
\begin{equation}
a_{I}^{\mathrm{int}}=\alpha _{I}\left( \left[ K^{A}\right] ,\left[ F^{a}%
\right] ,\left[ \Phi _{\Delta }^{\ast }\right] \right) \omega ^{I}\left(
\mathcal{F}_{\lambda \mu \nu \kappa }^{A},\mathcal{C}_{\mu \nu
}^{a},\partial _{\left[ \mu \right. }\mathcal{C}_{\left. \nu \kappa \right]
}^{a},C_{\nu }^{A}\right) .  \label{coltr79}
\end{equation}%
The notation $f([q])$ means that $f$ depends on $q$ and its derivatives up
to a finite order. In the above $\Phi _{\Delta }^{\ast }$ denote all the
antifields (see formula (\ref{antif})) and $\omega ^{I}$ represent the
elements of pure ghost number $I$ (and antighost number zero) of a basis in
the space of polynomials in $\mathcal{F}_{\lambda \mu \nu \kappa }^{A}$, $%
\mathcal{C}_{\mu \nu }^{a}$, $\partial _{\left[ \mu \right. }\mathcal{C}%
_{\left. \nu \kappa \right] }^{a}$, and $C_{\nu }^{A}$. The objects $\alpha
_{I}$ are non-trivial elements of the space $H^{0}\left( \gamma \right) $
and by hypothesis are polynomials in all the quantities on which they
depend, so they are nothing but the invariant polynomials of the free theory
(\ref{colrt1b}) in form degree equal to zero.

Replacing solution (\ref{coltr79}) into equation (\ref{coltv65d}), we get
that a necessary condition for the existence of non-trivial solutions $%
a_{I-1}^{\mathrm{int}}$ for $I>0$ is that the invariant polynomials $\alpha
_{I}$ appearing in (\ref{coltr79}) generate non-trivial elements from the
characteristic cohomology $H_{I}^{D}\left( \delta |d\right) $ in antighost
number $I>0$, maximum form degree, and pure ghost number equal to zero%
\footnote{\label{local copy(1)}We recall that the local cohomology $H_{\ast
}^{D}\left( \delta |d\right) $ is completely trivial at both strictly
positive antighost \textit{and} pure ghost numbers (for instance, see~\cite%
{gen1}, Theorem 5.4 and~\cite{commun1}).} computed in the algebra of local
forms, $\alpha _{I}d^{D}x\in H_{I}^{D}\left( \delta |d\right) $. As the free
model under study is a linear gauge theory of Cauchy order equal to four,
the general results from~\cite{gen1} ensure that
\begin{equation}
H_{j}^{D}\left( \delta |d\right) =0,\quad j>4.  \label{coltv83}
\end{equation}%
Meanwhile, it is possible to prove (see, for instance, Appendix B, Theorem
3, from~\cite{t31jhep}) that if $\alpha _{j}d^{D}x$ is a trivial element of $%
H_{j}^{D}\left( \delta |d\right) $ for $j>4$, then it can be chosen to be
trivial also in the local cohomology of the Koszul--Tate differential
computed in the space of invariant polynomials in antighost number $j$ and
maximum form degree (invariant characteristic cohomology), $H_{j}^{\mathrm{%
inv}D}\left( \delta |d\right) $. This is important since together with (\ref%
{coltv83}) ensures that the entire invariant characteristic cohomology is
trivial in antighost numbers strictly greater than four%
\begin{equation}
H_{j}^{\mathrm{inv}D}\left( \delta |d\right) =0,\quad j>4.  \label{coltv83b}
\end{equation}%
With the help of the general results from~\cite{PAUC09} on the
characteristic cohomology in the $(3,1)$ and respectively $(2,2)$ sector, we
identify the non-trivial and Poincar\'{e}-invariant representatives of the
spaces $\left( H_{j}^{D}\left( \delta |d\right) \right) _{j\geq 2}$ and $%
\left( H_{j}^{\mathrm{inv}D}\left( \delta |d\right) \right) _{j\geq 2}$.

\begin{table}[h]
\caption{Non-trivial representatives spanning $H_{j}^{D}\left( \delta |d\right) $ and
$\ H_{j}^{\mathrm{inv}D}\left( \delta |d\right) $}
\begin{center}
\begin{tabular}{@{}cc@{}}
\hline
agh & $H_{j}^{D}\left( \delta |d\right) $, $H_{j}^{\mathrm{inv}D}\left( \delta |d\right) $\\
\hline
$j>4$ & none \\
$j=4$ & $f_{\nu }^{A}C_{A}^{\ast \nu }d^{D}x$ \\
$j=3$ & $\left( f_{\nu \kappa }^{A}G_{A}^{\prime \ast \nu \kappa }+g_{\mu \nu
}^{a}\mathcal{C}_{a}^{\ast \mu \nu }\right) d^{D}x$ \\
$j=2$ & $\left( f_{\mu \nu \kappa }^{A}\mathcal{G}_{A}^{\prime \ast \mu \nu
|\kappa }+g_{\mu \nu \kappa }^{a}\mathcal{C}_{a}^{\ast \mu \nu |\kappa
}\right) d^{D}x$\\
\hline
\end{tabular}%
\end{center}
\label{coltvabledelta}
\end{table}%

All the coefficients from Table \ref{coltvabledelta} denoted by $f$ or $g$ define some constant,
non-derivative tensors. We remark that there is no non-trivial element in $\left( H_{j}^{D}\left(
\delta |d\right) \right) _{j\geq 2}$ or $\left( H_{j}^{\mathrm{inv}D}\left(
\delta |d\right) \right) _{j\geq 2}$ that effectively involves the curvature
tensors and/or their derivatives, and the same stands for the quantities
that are more than linear in the antifields and/or depend on their
derivatives. In principle, one can construct from the above elements in Table
\ref{coltvabledelta} other non-trivial invariant polynomials from $%
H_{j}^{D}\left( \delta |d\right) $ or $H_{j}^{\mathrm{inv}D}\left( \delta
|d\right) $, which depend on the space-time co-ordinates. For instance, it
can be checked by direct computation that $\mathcal{G}_{A}^{\prime \ast \mu
\nu |\kappa }f_{\mu \nu \kappa \rho }^{A}x^{\rho }d^{D}x$, with $f_{\mu \nu
\kappa \rho }^{A}$ some completely antisymmetric and constant tensors,
generate non-trivial representatives from both $H_{2}^{D}\left( \delta
|d\right) $ and $H_{2}^{\mathrm{inv}D}\left( \delta |d\right) $. However, we
will discard such candidates as they would break the Poincar\'{e} invariance
of the deformations. In contrast to the groups $\left( H_{j}^{D}\left(
\delta |d\right) \right) _{j\geq 2}$ and $\left( H_{j}^{\mathrm{inv}D}\left(
\delta |d\right) \right) _{j\geq 2}$, which are finite-dimensional, the
cohomology $H_{1}^{D}\left( \delta |d\right) $ at pure ghost number zero,
that is related to global symmetries and ordinary conservation laws, is
infinite-dimensional since the theory is free.

\section{First-order deformation\label{coldef1r22t31}}

The previous results on $H_{j}^{D}\left( \delta |d\right) $ and $H_{j}^{%
\mathrm{inv}D}\left( \delta |d\right) $ are important because they control
the obstructions to removing the antifields from the first-order
deformation. Indeed, due to (\ref{coltv83b}), it follows that we can
successively eliminate all the pieces with the antighost number $j>4$ from
the non-integrated density of the first-order deformation by adding only
trivial terms, so we can take, without loss of non-trivial objects, the
condition $I\leq 4$ in the first-order deformation. The last representative,
$a_{I}^{\mathrm{int}}$, is of the form (\ref{coltr79}), where the invariant
polynomials necessarily generate non-trivial objects from $H_{I}^{\mathrm{inv%
}D}\left( \delta |d\right) $ if $I=2,3,4$ and respectively from $%
H_{1}^{D}\left( \delta |d\right) $ if $I=1$. The cases $I=4$ and $I=3$ lead
to purely trivial solutions and will be analyzed in Appendix \ref{I=4,3cross}%
.

Next, we approach the case $I=2$%
\begin{equation}
a^{\mathrm{int}}=a_{0}^{\mathrm{int}}+a_{1}^{\mathrm{int}}+a_{2}^{\mathrm{int%
}},  \label{colrt57}
\end{equation}%
where $a_{2}^{\mathrm{int}}$ is the general solution to the homogeneous
equation $\gamma a_{2}^{\mathrm{int}}=0$, and thus of the type (\ref{coltr79}%
) for $I=2$, with $\alpha _{2}$ an invariant polynomial from $H_{2}^{\mathrm{%
inv}D}\left( \delta |d\right) $. With the help of Table \ref{coltvabledelta}
for $j=2$, we obtain that the general solution fulfilling all the working
hypotheses takes the form
\begin{equation}
a_{2}^{\mathrm{int}}=\mathcal{G}_{A}^{\prime \ast \mu \nu |\beta }\left(
P_{a\mu \nu \beta }^{A\lambda \rho }\mathcal{C}_{\lambda \rho }^{a}+Q_{a\mu
\nu \beta }^{A\lambda \rho \sigma }\partial _{\left[ \lambda \right. }%
\mathcal{C}_{\left. \rho \sigma \right] }^{a}\right) ,  \label{colrt58}
\end{equation}%
where $P_{a\mu \nu \beta }^{A\lambda \rho }$ and $Q_{a\mu \nu \beta
}^{A\lambda \rho \sigma }$ are some non-derivative, real constants, with the
properties $P_{a\mu \nu \beta }^{A\lambda \rho }=-P_{a\mu \nu \beta }^{A\rho
\lambda }$ and $Q_{a\mu \nu \beta }^{A\lambda \rho \sigma }=Q_{a\mu \nu
\beta }^{A\left[ \lambda \rho \sigma \right] }$. Acting with $\delta $ on (%
\ref{colrt58}), we infer%
\begin{equation}
\delta a_{2}^{\mathrm{int}}=\gamma \lambda _{1}+\partial ^{\mu }k_{\mu
}+t_{A}^{\ast \tau \mu \nu |\beta }P_{a\mu \nu \beta }^{A\lambda \rho
}\partial _{\left[ \tau \right. }\mathcal{C}_{\left. \lambda \rho \right]
}^{a},  \label{colrt59}
\end{equation}%
where%
\begin{equation}
\lambda _{1}=t_{A}^{\ast \tau \mu \nu |\beta }P_{a\mu \nu \beta }^{A\lambda
\rho }\mathcal{C}_{\lambda \rho |\tau }^{a}+\frac{3}{2}t_{A}^{\ast \tau \mu
\nu |\beta }Q_{a\mu \nu \beta }^{A\lambda \rho \sigma }\partial _{\lbrack
\lambda }\mathcal{C}_{\rho \sigma ]|\tau }^{a}.  \label{colrt60}
\end{equation}%
From (\ref{colrt59}) we find that $a_{1}^{\mathrm{int}}$ as solution to
equation (\ref{coltv65d}) for $I=2$ exists if and only if the last term in
the right-hand side of (\ref{colrt59}) is $\gamma $-exact modulo $d$%
\begin{equation}
t_{A}^{\ast \tau \mu \nu |\beta }P_{a\mu \nu \beta }^{A\lambda \rho
}\partial _{\left[ \tau \right. }\mathcal{C}_{\left. \lambda \rho \right]
}^{a}=\gamma u_{1}+\partial ^{\mu }q_{\mu }.  \label{colrt61}
\end{equation}%
Taking the (left) Euler--Lagrange derivative of the above equation with
respect to $t_{A}^{\ast \tau \mu \nu |\beta }$ and recalling the
anticommutativity of this operation with $\gamma $, we deduce%
\begin{equation}
P_{a\mu \nu \beta }^{A\lambda \rho }\partial _{\left[ \tau \right. }\mathcal{%
C}_{\left. \lambda \rho \right] }^{a}=\gamma \left( -\frac{\delta ^{L}u_{1}}{%
\delta t_{A}^{\ast \tau \mu \nu |\beta }}\right) .  \label{colrt62}
\end{equation}%
The previous equation reduces to the requirement that the object%
\begin{equation}
P_{a\mu \nu \beta }^{A\lambda \rho }\partial _{\left[ \tau \right. }\mathcal{%
C}_{\left. \lambda \rho \right] }^{a},  \label{colrt63}
\end{equation}%
which is a non-trivial element of $H^{2}\left( \gamma \right) $ (see
relation (\ref{coltr79})), must be $\gamma $-exact. This holds if and only
if $P_{a\mu \nu \beta }^{A\lambda \rho }=0$. The last result replaced in
formulas (\ref{colrt58})--(\ref{colrt60}) yields%
\begin{eqnarray}
a_{2}^{\mathrm{int}} &=&\mathcal{G}_{A}^{\prime \ast \mu \nu |\beta }Q_{a\mu
\nu \beta }^{A\lambda \rho \sigma }\partial _{\left[ \lambda \right. }%
\mathcal{C}_{\left. \rho \sigma \right] }^{a},  \label{colrt64} \\
\delta a_{2}^{\mathrm{int}} &=&\gamma \left( \frac{3}{2}t_{A}^{\ast \tau \mu
\nu |\beta }Q_{a\mu \nu \beta }^{A\lambda \rho \sigma }\partial _{\lbrack
\lambda }\mathcal{C}_{\rho \sigma ]|\tau }^{a}\right) +\partial ^{\mu
}k_{\mu }.  \label{colrt65}
\end{eqnarray}%
Equation (\ref{colrt65}) produces in a simple manner the solution $a_{1}^{%
\mathrm{int}}$ to equation (\ref{coltv65d}) for $I=2$ as
\begin{equation}
a_{1}^{\mathrm{int}}=-\frac{3}{2}t_{A}^{\ast \tau \mu \nu |\beta }Q_{a\mu
\nu \beta }^{A\lambda \rho \sigma }\partial _{\lbrack \lambda }\mathcal{C}%
_{\rho \sigma ]|\tau }^{a}+\bar{a}_{1}^{\mathrm{int}},  \label{colrt66}
\end{equation}%
where $\bar{a}_{1}^{\mathrm{int}}$ means the general solution to the
homogeneous equation $\gamma \bar{a}_{1}^{\mathrm{int}}=0$. Recalling once
more all the working hypotheses, we conclude that%
\begin{equation}
\bar{a}_{1}^{\mathrm{int}}=r_{a}^{\ast \mu \nu |\kappa \beta }Z_{A\mu \nu
\kappa \beta }^{a\sigma \tau \gamma \delta }\mathcal{F}_{\sigma \tau \gamma
\delta }^{A},  \label{colrt67}
\end{equation}%
where $Z_{A\mu \nu \kappa \beta }^{a\sigma \tau \gamma \delta }$ denote some
real, non-derivative constants, which are completely antisymmetric with
respect to the indices $\left\{ \sigma ,\tau ,\gamma ,\delta \right\} $. Due
to the mixed symmetry properties of the antifields $t_{A}^{\ast \tau \mu \nu
|\beta }$ and $r_{a}^{\ast \mu \nu |\kappa \beta }$, the only covariant
choice of the tensors $Q_{a\mu \nu \beta }^{A\lambda \rho \sigma }$ and $%
Z_{A\mu \nu \kappa \beta }^{a\sigma \tau \gamma \delta }$ in $D\geq 5$ that
does not end up with trivial solutions reads as%
\begin{equation}
Q_{a\mu \nu \beta }^{A\lambda \rho \sigma }=\frac{4}{3}f_{a}^{A}\varepsilon
_{\mu \nu \beta }^{\quad \ \ \lambda \rho \sigma }=\frac{4}{3}%
f_{a}^{A}\sigma ^{\lambda \lambda ^{\prime }}\sigma ^{\rho \rho ^{\prime
}}\sigma ^{\sigma \sigma ^{\prime }}\varepsilon _{\mu \nu \beta \lambda
^{\prime }\rho ^{\prime }\sigma ^{\prime }},\quad Z_{A\mu \nu \kappa \beta
}^{a\sigma \tau \gamma \delta }=0,  \label{colrt68}
\end{equation}%
with $\varepsilon _{\mu \nu \beta \lambda ^{\prime }\rho ^{\prime }\sigma
^{\prime }}$ the six-dimensional Levi--Civita symbol and $f_{a}^{A}$ some
real constants. Inserting (\ref{colrt68}) in formulas (\ref{colrt64}) and (%
\ref{colrt66})--(\ref{colrt67}) and recalling transformations (\ref{colt58a}%
), we finally obtain
\begin{gather}
a_{2}^{\mathrm{int}}=f_{a}^{A}\varepsilon ^{\lambda \mu \nu \kappa \beta
\gamma }\eta _{A\lambda \mu \nu }^{\ast }\partial _{\kappa }\mathcal{C}%
_{\beta \gamma }^{a},  \label{colrt69} \\
a_{1}^{\mathrm{int}}=-2f_{a}^{A}\varepsilon _{\lambda \mu \nu \rho \beta
\gamma }t_{A}^{\ast \lambda \mu \nu |\kappa }\left( \partial ^{\rho }%
\mathcal{C}_{\ \ \ \ \kappa }^{a\beta \gamma |}-\frac{1}{4}\delta _{\kappa
}^{\gamma }\partial ^{\left[ \rho \right. }\mathcal{C}_{\qquad \tau
}^{a\left. \beta \tau \right] |}\right) ,\quad \bar{a}_{1}^{\mathrm{int}}=0.
\label{colrt70}
\end{gather}%
The last term from the right-hand side of $a_{1}^{\mathrm{int}}$ is
vanishing due to the identity $t_{A}^{\ast \left[ \lambda \mu \nu |\kappa %
\right] }\equiv 0$, but it has been introduced in order to restore the mixed
symmetry $(3,1)$ of the Euler--Lagrange derivatives $\delta ^{L}a_{1}^{%
\mathrm{int}}/\delta t_{A}^{\ast \lambda \mu \nu |\kappa }$. By means of (%
\ref{colrt70}) we infer%
\begin{equation}
\delta a_{1}^{\mathrm{int}}=\gamma \left[ 2f_{a}^{A}\varepsilon ^{\lambda
\mu \nu \kappa \beta \gamma }t_{A\lambda \mu \nu |\rho }\left( \partial
_{\sigma }\partial _{\kappa }r_{\beta \gamma |}^{a\ \ \ \sigma \rho }-\frac{1%
}{2}\delta _{\gamma }^{\rho }\partial ^{\tau }\partial _{\kappa }r_{\beta
\tau }^{a}\right) \right] +\partial ^{\mu }p_{\mu }.  \label{colrt72}
\end{equation}%
The last relation generates the interacting Lagrangian at order one in the
coupling constant as the solution $a_{0}^{\mathrm{int}}$ of equation (\ref%
{coltv65e}) for $k=1$
\begin{equation}
a_{0}^{\mathrm{int}}=-2f_{a}^{A}\varepsilon ^{\lambda \mu \nu \kappa \beta
\gamma }t_{A\lambda \mu \nu |\rho }\left( \partial _{\sigma }\partial
_{\kappa }r_{\beta \gamma |}^{a\ \ \ \sigma \rho }-\frac{1}{2}\delta
_{\gamma }^{\rho }\partial ^{\tau }\partial _{\kappa }r_{\beta \tau
}^{a}\right) +\bar{a}_{0}^{\mathrm{int}}.  \label{colrt73}
\end{equation}%
Here, $\bar{a}_{0}^{\mathrm{int}}$ is the general solution to the
`homogeneous' equation
\begin{equation}
\gamma \bar{a}_{0}^{\mathrm{int}}=\partial _{\mu }\bar{m}_{\mathrm{int}%
}^{\mu },  \label{colrt71}
\end{equation}%
which cannot be replaced any longer with the homogeneous one since the
antighost number is vanishing, $I=0$. Without entering technical details, we
mention that the solution to equation (\ref{colrt71}) that fulfills all the working hypotheses
is also trivial
\begin{equation}
\bar{a}_{0}^{\mathrm{int}}=0.  \label{coltr102}
\end{equation}%
The proof of this result is done in Appendix \ref{aomcross}.

Putting together the results expressed by formulas (\ref{colrt69})--(\ref%
{colrt70}), (\ref{colrt73}), and (\ref{coltr102}), we can state that the
most general form of the first-order deformation associated with the free
theory (\ref{colrt1b}) reads%
\begin{eqnarray}
S_{1} &=&\int \left[ c_{a}r^{a}+f_{a}^{A}\varepsilon _{\mu \nu \kappa
\lambda \beta \gamma }\eta _{A}^{\ast \mu \nu \kappa }\partial ^{\lambda }%
\mathcal{C}^{a\beta \gamma }\right.  \notag \\
&&-2f_{a}^{A}\varepsilon _{\lambda \mu \nu \rho \beta \gamma }t_{A}^{\ast
\lambda \mu \nu |\kappa }\left( \partial ^{\rho }\mathcal{C}_{\quad \ \kappa
}^{a\beta \gamma |}-\frac{1}{4}\delta _{\kappa }^{\gamma }\partial ^{\left[
\rho \right. }\mathcal{C}_{\qquad \tau }^{a\left. \beta \tau \right]
|}\right)  \notag \\
&&\left. -2f_{a}^{A}\varepsilon ^{\lambda \mu \nu \kappa \beta \gamma
}t_{A\lambda \mu \nu |\rho }\left( \partial _{\sigma }\partial _{\kappa
}r_{\beta \gamma |}^{a\quad \sigma \rho }-\frac{1}{2}\delta _{\gamma }^{\rho
}\partial ^{\tau }\partial _{\kappa }r_{\beta \tau }^{a}\right) \right]
d^{6}x  \label{coltr103}
\end{eqnarray}%
and is defined on a space-time of dimension $D=6$.

\section{Higher-order deformations\label{coldefsupr22t31}}

In the sequel we approach the higher-order deformation equations. The
second-order deformation is controlled by equation (\ref{ec42}). After some
computations, with the help of relation (\ref{coltr103}) we arrive at%
\begin{equation}
\left( S_{1},S_{1}\right) =s\left[ f_{A}^{a}f_{b}^{A}\int \left(
10r_{a}^{\lambda \rho |\left[ \kappa \beta ,\gamma \right] }r_{\lambda \rho |%
\left[ \kappa \beta ,\gamma \right] }^{b}-12r_{a\lambda \rho |}^{\quad \left[
\kappa \beta ,\rho \right] }r_{\quad \ \left[ \kappa \beta ,\sigma \right]
}^{b\lambda \sigma |}\right) d^{6}x\right] ,  \label{colrt116}
\end{equation}%
such that the second-order deformation of the solution to the master
equation reduces to%
\begin{equation}
S_{2}=f_{A}^{a}f_{b}^{A}\int \left( -5r_{a}^{\lambda \rho |\left[ \kappa
\beta ,\gamma \right] }r_{\lambda \rho |\left[ \kappa \beta ,\gamma \right]
}^{b}+6r_{a\lambda \rho |}^{\quad \left[ \kappa \beta ,\rho \right]
}r_{\quad \ \left[ \kappa \beta ,\sigma \right] }^{b\lambda \sigma |}\right)
d^{6}x,  \label{coltr104}
\end{equation}%
where
\begin{equation}
r_{a}^{\lambda \rho |\left[ \kappa \beta ,\gamma \right] }=\partial ^{\gamma
}r_{a}^{\lambda \rho |\kappa \beta }+\partial ^{\beta }r_{a}^{\lambda \rho
|\gamma \kappa }+\partial ^{\kappa }r_{a}^{\lambda \rho |\beta \gamma }.
\label{colrt31}
\end{equation}%
Introducing relations (\ref{coltr103}) and (\ref{coltr104}) into the
equation corresponding to the third-order deformation, (\ref{ec43}), and
observing that $\left( S_{1},S_{2}\right) =0$, it follows that we can take%
\begin{equation}
S_{3}=0.  \label{colrt115}
\end{equation}%
Under these conditions, it is easy to see that all the remaining
higher-order deformation equations are fulfilled with the choice%
\begin{equation}
S_{k}=0,\quad k>3.  \label{colrt117}
\end{equation}%
In conclusion, the complete deformed solution to the master equation for the
model under study, which is consistent to all orders in the coupling
constant, reduces to%
\begin{equation}
\bar{S}=S+gS_{1}+g^{2}S_{2},  \label{colsbar2231}
\end{equation}%
where $S$ is the solution to the classical master equation for the free
model in $D=6$, (\ref{colrt2}), and $S_{1,2}$ are expressed by (\ref%
{coltr103}) and respectively (\ref{coltr104}).

\section{Identification of the coupled model\label{colmainres}}

From relations (\ref{colsbar2231}), (\ref{colrt2}), (\ref{coltr103}), and (%
\ref{coltr104}) we deduce the concrete form of the deformed solution to the
master equation%
\begin{eqnarray}
\bar{S} &=&S+g\int \left[ c_{a}r^{a}+f_{a}^{A}\varepsilon _{\mu \nu \kappa
\lambda \beta \gamma }\eta _{A}^{\ast \mu \nu \kappa }\partial ^{\lambda }%
\mathcal{C}^{a\beta \gamma }\right.  \notag \\
&&-2f_{a}^{A}\varepsilon _{\lambda \mu \nu \rho \beta \gamma }t_{A}^{\ast
\lambda \mu \nu |\kappa }\left( \partial ^{\rho }\mathcal{C}_{\quad \ \kappa
}^{a\beta \gamma |}-\frac{1}{4}\delta _{\kappa }^{\gamma }\partial ^{\left[
\rho \right. }\mathcal{C}_{\qquad \tau }^{a\left. \beta \tau \right]
|}\right)  \notag \\
&&\left. -2f_{a}^{A}\varepsilon ^{\lambda \mu \nu \kappa \beta \gamma
}t_{A\lambda \mu \nu |\rho }\left( \partial _{\sigma }\partial _{\kappa
}r_{\beta \gamma |}^{a\quad \sigma \rho }-\frac{1}{2}\delta _{\gamma }^{\rho
}\partial ^{\tau }\partial _{\kappa }r_{\beta \tau }^{a}\right) \right]
d^{6}x  \notag \\
&&-g^{2}\int f_{A}^{a}f_{b}^{A}\left( 5r_{a}^{\lambda \rho |\left[ \kappa
\beta ,\gamma \right] }r_{\lambda \rho |\left[ \kappa \beta ,\gamma \right]
}^{b}-6r_{a\lambda \rho |}^{\quad \left[ \kappa \beta ,\rho \right]
}r_{\quad \ \left[ \kappa \beta ,\sigma \right] }^{b\lambda \sigma |}\right)
d^{6}x.  \label{colrt30}
\end{eqnarray}%

The last formula enables us to identify the entire
information on the gauge structure of the interacting theory. In view of
this, we employ the fact that the piece of antighost number zero from $\bar{S%
}$ is nothing but the Lagrangian action of the coupled model, the terms of
antighost number one furnish the deformed gauge symmetries, and the
components of antighost number greater or equal to two offer us information
on the associated gauge algebra and the reducibility structure of the
generating set of deformed gauge transformations. As a consequence, we
deduce the coupled Lagrangian action%
\begin{eqnarray}
&&\bar{S}_{0}\left[ t_{\lambda \mu \nu |\kappa }^{A},r_{\mu \nu |\kappa
\beta }^{a}\right] =S_{0}\left[ t_{\lambda \mu \nu |\kappa }^{A},r_{\mu \nu
|\kappa \beta }^{a}\right]  \notag \\
&&+g\int \left[ c_{a}r^{a}-2f_{a}^{A}\varepsilon ^{\lambda \mu \nu \kappa
\beta \gamma }t_{A\lambda \mu \nu |\rho }\left( \partial _{\sigma }\partial
_{\kappa }r_{\beta \gamma |}^{a\quad \sigma \rho }-\frac{1}{2}\delta
_{\gamma }^{\rho }\partial ^{\tau }\partial _{\kappa }r_{\beta \tau
}^{a}\right) \right.  \notag \\
&&\left. -gf_{A}^{a}f_{b}^{A}\left( 5r_{a}^{\lambda \rho |\left[ \kappa
\beta ,\gamma \right] }r_{\lambda \rho |\left[ \kappa \beta ,\gamma \right]
}^{b}-6r_{a\lambda \rho |}^{\quad \left[ \kappa \beta ,\rho \right]
}r_{\quad \ \left[ \kappa \beta ,\sigma \right] }^{b\lambda \sigma |}\right) %
\right] d^{6}x,  \label{colrt32}
\end{eqnarray}%
where $S_{0}\left[ t_{\lambda \mu \nu |\kappa }^{A},r_{\mu \nu |\kappa \beta
}^{a}\right] $ is the free action (\ref{colrt1b}) in $D=6$ space-time
dimensions. We observe that action (\ref{colrt32}) contains only
mixing-component terms of order one and two in the coupling constant.
Apparently, it seems that (\ref{colrt32}) contains non-trivial couplings
between different tensor fields with the mixed symmetry of the Riemann tensor%
\begin{equation}
-g^{2}f_{A}^{a}f_{b}^{A}\left( 5r_{a}^{\lambda \rho |\left[ \kappa \beta
,\gamma \right] }r_{\lambda \rho |\left[ \kappa \beta ,\gamma \right]
}^{b}-6r_{a\lambda \rho |}^{\quad \left[ \kappa \beta ,\rho \right]
}r_{\quad \ \left[ \kappa \beta ,\sigma \right] }^{b\lambda \sigma |}\right)
,\quad a\neq b.  \label{colrt32a}
\end{equation}%
The appearance of these cross-couplings is dictated by the properties of the
matrix $M$ of elements $M_{b}^{a}=f_{A}^{a}f_{b}^{A}$.

Let us analyze the properties of the quadratic matrix $M$. It is more
convenient to work with the symmetric matrix $\hat{M}=\left( M_{ab}\right) $%
, of elements $M_{ab}=f_{a}^{A}f_{b}^{B}k_{AB}$. From (\ref{colr1}) and (\ref%
{colrt32}) we observe that there appear effective cross-couplings among
different fields from the collection $\left\{ r_{\mu \nu |\kappa \beta
}^{a}\right\} _{a=\overline{1,n}}$ if and only if the symmetric matrices $%
\hat{M}=\left( M_{ab}\right) $ and $\hat{k}=\left( k_{ab}\right) $ are
simultaneously diagonalizable. We recall $\hat{k}$ is the quadratic form
defined by the kinetic terms of action (\ref{colr1}), or, in other words,
the metric tensor in the inner space of collection indices $a=\overline{1,n}$%
. This means that there exists an orthogonal matrix $\hat{O}=\left( O_{\ \
b}^{a}\right) $ that diagonalizes simultaneously~\cite{algebra} $\hat{M}$\
and $\hat{k}$, i.e.%
\begin{equation}
O_{\ \ a}^{c}O_{\ \ b}^{d}k_{cd}=k_{a}\delta _{ab},\quad O_{\ \ a}^{c}O_{\ \
b}^{d}M_{cd}=m_{a}\delta _{ab},  \label{ort1}
\end{equation}%
where $k_{a}$ represent the eigenvalues of the matrix $\hat{k}$ and $m_{a}$
those of $\hat{M}$. Indeed, if there exists a matrix $\hat{O}$ that
satisfies the conditions (\ref{ort1}), then action (\ref{colrt32}) can be
brought to the form%
\begin{eqnarray}
&&\bar{S}_{0}\left[ t_{\lambda \mu \nu |\kappa }^{A},r_{\mu \nu |\kappa
\beta }^{a}\right] =\bar{S}_{0}^{\prime }\left[ t_{\lambda \mu \nu |\kappa
}^{A},r_{\mu \nu |\kappa \beta }^{\prime a}\right] =S_{0}^{\mathrm{t}}\left[
t_{\lambda \mu \nu |\kappa }^{A}\right]  \notag \\
&&+\int \sum\limits_{a=1}^{n}k_{a}\left\{ -\frac{1}{2}\left[ \left( \partial
_{\mu }r^{\prime a\mu \nu |\kappa \beta }\right) \left( \partial ^{\lambda
}r_{\lambda \nu |\kappa \beta }^{\prime a}\right) +\left( \partial ^{\lambda
}r^{\prime a\nu \beta }\right) \left( \partial _{\lambda }r_{\nu \beta
}^{\prime a}\right) \right. \right.  \notag \\
&&\left. +\left( \partial _{\nu }r^{\prime a\nu \beta }\right) \left(
\partial _{\beta }r^{\prime a}\right) \right] +\frac{1}{8}\left[ \left(
\partial ^{\lambda }r^{\prime a\mu \nu |\kappa \beta }\right) \left(
\partial _{\lambda }r_{\mu \nu |\kappa \beta }^{\prime a}\right) +\left(
\partial ^{\lambda }r^{\prime a}\right) \left( \partial _{\lambda }r^{\prime
a}\right) \right]  \notag \\
&&\left. -\left( \partial _{\mu }r^{\prime a\mu \nu |\kappa \beta }\right)
\left( \partial _{\beta }r_{\nu \kappa }^{\prime a}\right) +\left( \partial
_{\nu }r^{\prime a\nu \beta }\right) \left( \partial ^{\lambda }r_{\lambda
\beta }^{\prime a}\right) \right\} d^{6}x  \notag \\
&&+g\int \left[ c_{a}^{\prime }r^{\prime a}-2f_{a}^{\prime A}\varepsilon
^{\lambda \mu \nu \kappa \beta \gamma }t_{A\lambda \mu \nu |\rho }\left(
\partial _{\sigma }\partial _{\kappa }r_{\beta \gamma |}^{\prime a\quad
\sigma \rho }-\frac{1}{2}\delta _{\gamma }^{\rho }\partial ^{\tau }\partial
_{\kappa }r_{\beta \tau }^{\prime a}\right) \right.  \notag \\
&&\left. -g\sum\limits_{a=1}^{n}m_{a}\left( 5r^{\prime a\lambda \rho |\left[
\kappa \beta ,\gamma \right] }r_{\lambda \rho |\left[ \kappa \beta ,\gamma %
\right] }^{\prime a}-6r_{\lambda \rho |}^{\prime a\quad \left[ \kappa \beta
,\rho \right] }r_{\quad \ \left[ \kappa \beta ,\sigma \right] }^{\prime
a\lambda \sigma |}\right) \right] d^{6}x,  \label{ort2}
\end{eqnarray}%
where we made the transformations%
\begin{equation}
r_{\mu \nu |\kappa \beta }^{a}\rightarrow r_{\mu \nu |\kappa \beta }^{\prime
a}=\bar{O}_{\ \ b}^{a}r_{\mu \nu |\kappa \beta }^{b},  \label{ort3}
\end{equation}%
and used the notations%
\begin{equation}
c_{a}^{\prime }=c_{b}O_{\ \ a}^{b},\quad f_{a}^{\prime A}=f_{b}^{A}O_{\ \
a}^{b}.  \label{ort4}
\end{equation}%
The quantities $\bar{O}_{\ \ b}^{a}$ from (\ref{ort3}) denote the elements
of the inverse of $\hat{O}$. These considerations allow us to conclude that:

\begin{enumerate}
\item If the matrix $\hat{k}$ is positive-definite, then the symmetric
matrices $\hat{M}=\left( M_{ab}\right) $ and $\hat{k}=\left( k_{ab}\right) $
are simultaneously diagonalizable and hence there appear no cross-couplings
among different fields from the collection $\left\{ r_{\mu \nu |\kappa \beta
}^{a}\right\} _{a=\overline{1,n}}$. Taking $\hat{k}$ to be positive-definite
might be essential for the physical consistency of the theory (absence of
negative-energy excitations or stability of the Minkowski vacuum);

\item If the matrix $\hat{k}$ is indefinite, then the matrices $\hat{M}$\
and $\hat{k}$ cannot be diagonalized simultaneously (because then the matrix
$\hat{C}=\hat{k}^{-1}\hat{M}$\ is not normal~\cite{algebra}) and therefore
there appear cross-couplings among different fields from the collection $%
\left\{ r_{\mu \nu |\kappa \beta }^{a}\right\} _{a=\overline{1,n}}$.
\end{enumerate}

The terms from (\ref{colrt30}) that are linear in the antifields of the
original fields give the gauge transformations of the deformed Lagrangian
action, (\ref{colrt32}), by replacing the ghosts with the corresponding
gauge parameters%
\begin{eqnarray}
\bar{\delta}_{\epsilon ,\chi ,\xi }t_{\lambda \mu \nu |\kappa }^{A}
&=&3\partial _{\kappa }\epsilon _{\lambda \mu \nu }^{A}+\partial _{\left[
\lambda \right. }\epsilon _{\left. \mu \nu \right] \kappa }^{A}+\partial _{%
\left[ \lambda \right. }\chi _{\left. \mu \nu \right] |\kappa }^{A}  \notag
\\
&&-2gf_{a}^{A}\varepsilon _{\lambda \mu \nu \rho \beta \gamma }\left(
\partial ^{\rho }\xi _{\quad \ \kappa }^{a\beta \gamma |}-\frac{1}{4}\delta
_{\kappa }^{\gamma }\partial ^{\left[ \rho \right. }\xi _{\qquad \tau
}^{a\left. \beta \tau \right] |}\right) ,  \label{colrt33}
\end{eqnarray}%
\begin{equation}
\bar{\delta}_{\xi }r_{\mu \nu |\kappa \beta }^{a}=\partial _{\mu }\xi
_{\kappa \beta |\nu }^{a}-\partial _{\nu }\xi _{\kappa \beta |\mu
}^{a}+\partial _{\kappa }\xi _{\mu \nu |\beta }^{a}-\partial _{\beta }\xi
_{\mu \nu |\kappa }^{a}=\delta _{\xi }r_{\mu \nu |\kappa \beta }^{a}.
\label{colrt34}
\end{equation}%
It is interesting to note that only the gauge transformations of the tensor
fields $(3,1)$ are modified during the deformation process. This is enforced
at order one in the coupling constant by terms linear in the first-order
derivatives of the gauge parameters from the $(2,2)$ sector. From the terms
of antighost number equal to two present in (\ref{colrt30}) we learn that
only the first-order reducibility functions are modified at order one in the
coupling constant, the others coinciding with the original ones.
Consequently, the first-order reducibility relations corresponding to the
fields $t_{\lambda \mu \nu |\kappa }^{A}$ take place off-shell, like the
free ones, while the first-order reducibility relations associated with the
fields $r_{\mu \nu |\kappa \beta }^{a}$ remain the original ones. Since
there are no other terms of antighost number two in (\ref{colrt30}), it
follows that the gauge algebra of the coupled model is unchanged by the
deformation procedure, being the same Abelian one like for the starting free
theory. The structure of pieces with the antighost number equal to three
from (\ref{colrt30}) implies that the second-order reducibility functions
remain the same, and hence the second-order reducibility relations are
exactly the initial ones. It is easy to see from (\ref{colrt32})--(\ref%
{colrt34}) that if we impose the PT-invariance at the level of the coupled
model, then we obtain no interactions at all.

It is important to stress that the problem of obtaining consistent
interactions strongly depends on the space-time dimension. For instance, if
one starts with action (\ref{colrt1b}) in $D>6$, then one inexorably gets $%
\bar{S}=S+g\int c_{a}r^{a}d^{D}x$, so no cross-interaction term can be added
to either the original Lagrangian or its gauge transformations.

\section{Conclusions}

Results (\ref{colrt30})--(\ref{colrt34}) lead to the following main result
of our work:\textit{\ }under the hypotheses of analyticity of deformations
in the coupling constant, space-time locality, Lorentz covariance, and
Poincar\'{e} invariance, combined with the requirement that the interaction
vertices contain at most two space-time derivatives of the fields, there
appear consistent cross-couplings in $D=6$ between a collection of massless
tensor fields with the mixed symmetry $(3,1)$ and a collection of massless
tensor fields with the mixed symmetry of the Riemann tensor, with the
property that they modify the free action and its gauge symmetries.\ The
existence of cross-couplings among different fields with the mixed symmetry
of the Riemann tensor is essentially dictated by the behaviour of the metric
tensor in the inner space of collection indices $a=\overline{1,n}$, $\hat{k}%
=\left( k_{ab}\right) $. Thus, if $\hat{k}$ is positive-definite, then there
appear no cross-couplings among different fields with the mixed symmetry of
the Riemann tensor. On the contrary, if $\hat{k}$ is indefinite, then there
are allowed cross-couplings among different fields from this collection.

\section*{Acknowledgments}

One of the authors (E.M.B.) acknowledges financial support from the contract
464/2009 in the framework of the programme IDEI of C.N.C.S.I.S. (Romanian
National Council for Academic Scientific Research).

\appendix

\section{Proof of the triviality of the first-order deformation
for $I=4$ and $I=3$\label{I=4,3cross}}

In order to solve the third equation from (\ref{coltv85c}), we decompose $a^{%
\mathrm{int}}$ along the antighost number and stop at $I=4$%
\begin{equation}
a^{\mathrm{int}}=a_{0}^{\mathrm{int}}+a_{1}^{\mathrm{int}}+a_{2}^{\mathrm{int%
}}+a_{3}^{\mathrm{int}}+a_{4}^{\mathrm{int}},  \label{tv87}
\end{equation}%
where $a_{4}^{\mathrm{int}}$ can be taken as solution to the equation $%
\gamma a_{4}^{\mathrm{int}}=0$, and therefore it is of the form (\ref%
{coltr79}) for $I=4$, with $\alpha _{4}d^{D}x$ an invariant polynomial from $%
H_{4}^{\mathrm{inv}D}\left( \delta |d\right) $. Because $H_{4}^{\mathrm{inv}%
D}\left( \delta |d\right) $ is spanned by $C_{A}^{\ast \mu }$ (see Table
\ref{coltvabledelta}) and $a_{4}^{\mathrm{int}}$ must yield cross-couplings
between $t_{\lambda \mu \nu |\kappa }^{A}$ and $r_{\mu \nu |\kappa \beta
}^{a}$ with maximum two space-time derivatives, it follows that the eligible
basis elements at pure ghost number equal to four remain%
\begin{equation}
\omega ^{4}:\left( \mathcal{C}_{\kappa \beta }^{a}\mathcal{C}_{\lambda \rho
}^{b},\mathcal{C}_{\kappa \beta }^{a}\partial _{\left[ \lambda \right. }%
\mathcal{C}_{\left. \rho \sigma \right] }^{b}\right) .  \label{colrt38}
\end{equation}%
So, up to trivial, $\gamma $-exact contributions, we have that%
\begin{equation}
a_{4}^{\mathrm{int}}=C_{A}^{\ast \mu }\left( M_{ab\mu }^{A\kappa \beta
\lambda \rho }\mathcal{C}_{\kappa \beta }^{a}\mathcal{C}_{\lambda \rho
}^{b}+N_{ab\mu }^{A\kappa \beta \lambda \rho \sigma }\mathcal{C}_{\kappa
\beta }^{a}\partial _{\left[ \lambda \right. }\mathcal{C}_{\left. \rho
\sigma \right] }^{b}\right) ,  \label{coltv88}
\end{equation}%
where $M_{ab\mu }^{A\kappa \beta \lambda \rho }=-M_{ab\mu }^{A\beta \kappa
\lambda \rho }=-M_{ab\mu }^{A\kappa \beta \rho \lambda }=M_{ba\mu
}^{A\lambda \rho \kappa \beta }$ and $N_{ab\mu }^{A\kappa \beta \lambda \rho
\sigma }=N_{ab\mu }^{A\left[ \kappa \beta \right] \lambda \rho \sigma
}=N_{ab\mu }^{A\kappa \beta \lbrack \lambda \rho \sigma ]}$ are some
non-derivative, real constants. Replacing $a_{4}^{\mathrm{int}}$ into an
equation similar to (\ref{coltv65d}) for $I=4$ and computing $\delta a_{4}^{%
\mathrm{int}}$, it follows that%
\begin{equation}
\delta a_{4}^{\mathrm{int}}=\gamma \lambda _{3}+\partial ^{\mu }\tau _{\mu
}-2G_{A}^{\prime \ast \nu \mu }\partial _{\left[ \nu \right. }\mathcal{C}%
_{\left. \kappa \beta \right] }^{a}\left( 2M_{ab\mu }^{A\kappa \beta \lambda
\rho }\mathcal{C}_{\lambda \rho }^{b}+N_{ab\mu }^{A\kappa \beta \lambda \rho
\sigma }\partial _{\left[ \lambda \right. }\mathcal{C}_{\left. \rho \sigma %
\right] }^{b}\right) ,  \label{coltv89}
\end{equation}%
where%
\begin{eqnarray}
\lambda _{3} &=&-G_{A}^{\prime \ast \nu \mu }\left[ 2\mathcal{C}_{\kappa
\beta |\nu }^{a}\left( 2M_{ab\mu }^{A\kappa \beta \lambda \rho }\mathcal{C}%
_{\lambda \rho }^{b}+N_{ab\mu }^{A\kappa \beta \lambda \rho \sigma }\partial
_{\left[ \lambda \right. }\mathcal{C}_{\left. \rho \sigma \right]
}^{b}\right) \right.  \notag \\
&&\left. +3\mathcal{C}_{\kappa \beta }^{a}N_{ab\mu }^{A\kappa \beta \lambda
\rho \sigma }\partial _{\lbrack \lambda }\mathcal{C}_{\rho \sigma ]|\nu }^{b}%
\right] .  \label{colrt39}
\end{eqnarray}%
Thus, $a_{3}^{\mathrm{int}}$ exists if and only if the third term in the
right-hand side of (\ref{coltv89}) can be written in a $\gamma $-exact
modulo $d$ form%
\begin{equation}
G_{A}^{\prime \ast \nu \mu }\partial _{\left[ \nu \right. }\mathcal{C}%
_{\left. \kappa \beta \right] }^{a}\left( 2M_{ab\mu }^{A\kappa \beta \lambda
\rho }\mathcal{C}_{\lambda \rho }^{b}+N_{ab\mu }^{A\kappa \beta \lambda \rho
\sigma }\partial _{\left[ \lambda \right. }\mathcal{C}_{\left. \rho \sigma %
\right] }^{b}\right) =\gamma u_{3}+\partial ^{\mu }\pi _{\mu }.
\label{colrt40}
\end{equation}%
Taking the (left) Euler--Lagrange derivative of the above equation with
respect to $G_{A}^{\prime \ast \nu \mu }$ and recalling the
anticommutativity of this operation with $\gamma $, we obtain%
\begin{equation}
\partial _{\left[ \nu \right. }\mathcal{C}_{\left. \kappa \beta \right]
}^{a}\left( 2M_{ab\mu }^{A\kappa \beta \lambda \rho }\mathcal{C}_{\lambda
\rho }^{b}+N_{ab\mu }^{A\kappa \beta \lambda \rho \sigma }\partial _{\left[
\lambda \right. }\mathcal{C}_{\left. \rho \sigma \right] }^{b}\right)
=\gamma \left( -\frac{\delta ^{L}u_{3}}{\delta G_{A}^{\prime \ast \nu \mu }}%
\right) .  \label{colrt41}
\end{equation}%
The last relation shows that the object%
\begin{equation}
\partial _{\left[ \nu \right. }\mathcal{C}_{\left. \kappa \beta \right]
}^{a}\left( 2M_{ab\mu }^{A\kappa \beta \lambda \rho }\mathcal{C}_{\lambda
\rho }^{b}+N_{ab\mu }^{A\kappa \beta \lambda \rho \sigma }\partial _{\left[
\lambda \right. }\mathcal{C}_{\left. \rho \sigma \right] }^{b}\right) ,
\label{colrt42}
\end{equation}%
which is a non-trivial element of $H^{4}\left( \gamma \right) $ (see formula
(\ref{coltr79})), must be $\gamma $-exact. This takes place if and only if $%
M_{ab\mu }^{A\kappa \beta \lambda \rho }=0=N_{ab\mu }^{A\kappa \beta \lambda
\rho \sigma }$, which further implies%
\begin{equation}
a_{4}^{\mathrm{int}}=0,  \label{colrt43}
\end{equation}%
and hence the first-order deformation in the cross-coupling sector cannot
end non-trivially at antighost number $I=4$.

The case $I=3$ is solved in a similar manner and leads to the result $a_{3}^{%
\mathrm{int}}=0$.

\section{Proof of the result (\protect\ref{coltr102})\label{aomcross}}

Next, we investigate the solutions to (\ref{colrt71}). There are two main
types of solutions to this equation. The first type, to be denoted by $\bar{a%
}_{0}^{\prime \mathrm{int}}$, corresponds to $\bar{m}_{\mathrm{int}}^{\mu
}=0 $ and is given by gauge-invariant, non-integrated densities constructed
out of the original fields and their space-time derivatives, which,
according to (\ref{coltr79}), are of the form $\bar{a}_{0}^{\prime \mathrm{%
int}}=\bar{a}_{0}^{\prime \mathrm{int}}\left( \left[ K_{\lambda \mu \nu \xi
|\kappa \beta }^{A}\right] ,\left[ F_{\mu \nu \lambda |\kappa \beta \gamma
}^{a}\right] \right) $, up to the condition that they effectively describe
cross-couplings between the two types of fields and cannot be written in a
divergence-like form. Such a solution implies at least four derivatives of
the fields and consequently must be forbidden by setting $\bar{a}%
_{0}^{\prime \mathrm{int}}=0$.

The second kind of solutions is associated with $\bar{m}_{\mathrm{int}}^{\mu
}\neq 0$ in (\ref{colrt71}), being understood that we discard the
divergence-like quantities and maintain the condition on the maximum
derivative order of the interacting Lagrangian being equal to two. In order
to solve this equation we start from the requirement that $\bar{a}_{0}^{%
\mathrm{int}}$ may contain at most two derivatives, so it can be decomposed
like
\begin{equation}
\bar{a}_{0}^{\mathrm{int}}=\omega _{0}+\omega _{1}+\omega _{2},
\label{coltww60}
\end{equation}%
where $\left( \omega _{i}\right) _{i=\overline{0,2}}$ contains $i$
derivatives. Due to the different number of derivatives in the components $%
\omega _{0}$, $\omega _{1}$, and $\omega _{2}$, equation (\ref{colrt71}) is
equivalent to three independent equations
\begin{equation}
\gamma \omega _{k}=\partial _{\mu }j_{k}^{\mu },\quad k=0,1,2.
\label{coltwwxy}
\end{equation}

Equation (\ref{coltwwxy}) for $k=0$ implies the (necessary) conditions
\begin{equation}
\partial _{\lambda }\left( \frac{\partial \omega _{0}}{\partial t_{\lambda
\mu \nu |\kappa }^{A}}\right) =0,\quad \partial _{\kappa }\left( \frac{%
\partial \omega _{0}}{\partial t_{\lambda \mu \nu |\kappa }^{A}}\right)
=0,\quad \partial _{\mu }\left( \frac{\partial \omega _{0}}{\partial r_{\mu
\nu |\kappa \beta }^{a}}\right) =0.  \label{coltvcond0}
\end{equation}%
The last equation from (\ref{coltvcond0}) possesses only the constant
solution
\begin{equation}
\frac{\partial \omega _{0}}{\partial r_{\mu \nu |\kappa \beta }^{a}}%
=k_{a}\left( \sigma ^{\mu \kappa }\sigma ^{\nu \beta }-\sigma ^{\mu \beta
}\sigma ^{\nu \kappa }\right) ,  \label{coltvsol0}
\end{equation}%
where $k_{a}$ are some real constants, so we find that%
\begin{equation}
\omega _{0}=2k_{a}r^{a}+B\left( t_{\lambda \mu \nu |\kappa }^{A}\right) .
\label{colrt75}
\end{equation}%
Since $\omega _{0}$ provides no cross-couplings between $t_{\lambda \mu \nu
|\kappa }^{A}$ and $r_{\mu \nu |\kappa \beta }^{a}$, we can take
\begin{equation}
\omega _{0}=0  \label{colomega0}
\end{equation}%
in (\ref{coltww60}).

As a digression, we note that the general solution to the equations
\begin{equation}
\partial _{\lambda }\bar{T}_{A}^{\lambda \mu \nu |\kappa }=0,\quad \partial
_{\kappa }\bar{T}_{A}^{\lambda \mu \nu |\kappa }=0  \label{coltv32c}
\end{equation}%
(with $\bar{T}_{A}^{\lambda \mu \nu |\kappa }$ some covariant tensor fields
with the mixed symmetry $(3,1)$) reads as~\cite{t31jhep}
\begin{equation}
\bar{T}_{A}^{\lambda \mu \nu |\kappa }=\partial _{\xi }\partial _{\beta }%
\bar{\Phi}_{A}^{\lambda \mu \nu \xi |\kappa \beta },  \label{coltv32d}
\end{equation}%
where $\bar{\Phi}_{A}^{\lambda \mu \nu \xi |\kappa \beta }$ are some
tensors with the mixed symmetry $(4,2)$. A constant solution $C_{A}^{\lambda
\mu \nu |\kappa }$ is excluded from covariance arguments due to the mixed
symmetry $(3,1)$. Along the same line, the general solution to the equations
\begin{equation}
\partial _{\mu }\bar{R}_{a}^{\mu \nu |\kappa \beta }=0  \label{colr40a}
\end{equation}%
(with $\bar{R}_{a}^{\mu \nu |\kappa \beta }$ some covariant tensor fields
with the mixed symmetry $(2,2)$) is represented by~\cite{r22}
\begin{equation}
\bar{R}_{a}^{\mu \nu |\kappa \beta }=\partial _{\rho }\partial _{\gamma }%
\bar{\Omega}_{a}^{\mu \nu \rho |\kappa \beta \gamma }+k_{a}\left( \sigma
^{\mu \kappa }\sigma ^{\nu \beta }-\sigma ^{\mu \beta }\sigma ^{\nu \kappa
}\right) ,  \label{colr40b}
\end{equation}%
where $\bar{\Omega}_{a}^{\mu \nu \rho |\kappa \beta \gamma }$ are some
tensors with the mixed symmetry $(3,3)$ and $k_{a}$ some arbitrary, real
constants. Now, it is clear why the solution to the last equation from (\ref%
{coltvcond0}) reduces to (\ref{coltvsol0}): $\partial \omega _{0}/\partial
r_{\mu \nu |\kappa \beta }^{a}$ display the mixed symmetry $(2,2)$, but are
derivative-free by assumption, so some terms similar to the former ones from
the right-hand side of (\ref{colr40b}) are forbidden.

Equation (\ref{coltwwxy}) for $k=1$ leads to the requirements
\begin{equation}
\partial _{\lambda }\left( \frac{\delta \omega _{1}}{\delta t_{\lambda \mu
\nu |\kappa }^{A}}\right) =0,\quad \partial _{\kappa }\left( \frac{\delta
\omega _{1}}{\delta t_{\lambda \mu \nu |\kappa }^{A}}\right) =0,\quad
\partial _{\mu }\left( \frac{\delta \omega _{1}}{\delta r_{\mu \nu |\kappa
\beta }^{a}}\right) =0,  \label{coltvcond1}
\end{equation}%
where $\delta \omega _{1}/\delta t_{\lambda \mu \nu |\kappa }^{A}$ and $%
\delta \omega _{1}/\delta r_{\mu \nu |\kappa \beta }^{a}$ denote the
Euler--Lagrange derivatives of $\omega _{1}$ with respect to the
corresponding fields. Looking at (\ref{coltv32d}) and (\ref{colr40b}) and
recalling that $\omega _{1}$ is by hypothesis of order one in the space-time
derivatives of the fields, the only solution to equations (\ref{coltvcond1})
reduces to%
\begin{equation}
\frac{\delta \omega _{1}}{\delta r_{\mu \nu |\kappa \beta }^{a}}=0=\frac{%
\delta \omega _{1}}{\delta t_{\lambda \mu \nu |\kappa }^{A}}.
\label{coltvsol1}
\end{equation}%
This solution forbids the cross-couplings between the two types of fields,
so we can safely take
\begin{equation}
\omega _{1}=0.  \label{colomega1}
\end{equation}

Finally, we pass to equation (\ref{coltwwxy}) for $k=2$, which produces the
restrictions
\begin{equation}
\partial _{\lambda }\left( \frac{\delta \omega _{2}}{\delta t_{\lambda \mu
\nu |\kappa }^{A}}\right) =0,\quad \partial _{\kappa }\left( \frac{\delta
\omega _{2}}{\delta t_{\lambda \mu \nu |\kappa }^{A}}\right) =0,\quad
\partial _{\mu }\left( \frac{\delta \omega _{2}}{\delta r_{\mu \nu |\kappa
\beta }^{a}}\right) =0,  \label{coltvcond2}
\end{equation}%
with the solutions (see formulas (\ref{coltv32d}) and (\ref{colr40b}))%
\begin{equation}
\frac{\delta \omega _{2}}{\delta t_{\lambda \mu \nu |\kappa }^{A}}=\partial
_{\gamma }\partial _{\sigma }W_{A}^{\lambda \mu \nu \gamma |\kappa \sigma
},\quad \frac{\delta \omega _{2}}{\delta r_{\mu \nu |\kappa \beta }^{a}}%
=\partial _{\gamma }\partial _{\sigma }U_{a}^{\mu \nu \gamma |\kappa \beta
\sigma }.  \label{coltvsol2}
\end{equation}%
The tensors $W_{A}^{\lambda \mu \nu \gamma |\kappa \sigma }$ have the mixed
symmetry of the curvature tensors $K_{A}^{\lambda \mu \nu \gamma |\kappa
\sigma }$ and the tensors $U_{a}^{\mu \nu \gamma |\kappa \beta \sigma }$
exhibit the mixed symmetry of the curvature tensors $F_{a}^{\mu \nu \gamma
|\kappa \beta \sigma }$. Both types of tensors are derivative-free since $%
\omega _{2}$ contains precisely two derivatives of the fields. At this stage
it is useful to introduce a derivation in the algebra of the fields and of
their derivatives that counts the powers of the fields and of their
derivatives
\begin{eqnarray}
N &=&\sum\limits_{k\geq 0}\left[ \left( \partial _{\mu _{1}\ldots \mu
_{k}}t_{\lambda \mu \nu |\kappa }^{A}\right) \frac{\partial }{\partial
\left( \partial _{\mu _{1}\ldots \mu _{k}}t_{\lambda \mu \nu |\kappa
}^{A}\right) }\right.  \notag \\
&&\left. +\left( \partial _{\mu _{1}\ldots \mu _{k}}r_{\mu \nu |\kappa \beta
}^{a}\right) \frac{\partial }{\partial \left( \partial _{\mu _{1}\ldots \mu
_{k}}r_{\mu \nu |\kappa \beta }^{a}\right) }\right] ,  \label{coltww74}
\end{eqnarray}%
so for every non-integrated density $\rho $ we have that
\begin{equation}
N\rho =t_{\lambda \mu \nu |\kappa }^{A}\frac{\delta \rho }{\delta t_{\lambda
\mu \nu |\kappa }^{A}}+r_{\mu \nu |\kappa \beta }^{a}\frac{\delta \rho }{%
\delta r_{\mu \nu |\kappa \beta }^{a}}+\partial _{\mu }s^{\mu },
\label{coltww75}
\end{equation}%
where $\delta \rho /\delta t_{\mu \nu |\kappa \beta }^{A}$ and $\delta \rho
/\delta r_{\mu \nu |\kappa \beta }^{a}$ denote the variational derivatives
of $\rho $ with respect to the fields. If $\rho ^{\left( l\right) }$ is a
homogeneous polynomial of order $l>0$ in the fields $\left\{ t_{\lambda \mu
\nu |\kappa }^{A},r_{\mu \nu |\kappa \beta }^{a}\right\} $ and their
derivatives, then $N\rho ^{\left( l\right) }=l\rho ^{\left( l\right) }$.
Using (\ref{coltvsol2}) and (\ref{coltww75}), we find that
\begin{equation}
N\omega _{2}=\frac{1}{8}K_{\lambda \mu \nu \gamma |\kappa \sigma
}^{A}W_{A}^{\lambda \mu \nu \gamma |\kappa \sigma }+\frac{1}{9}F_{\mu \nu
\gamma |\kappa \beta \sigma }^{a}U_{a}^{\mu \nu \gamma |\kappa \beta \sigma
}+\partial _{\mu }v^{\mu }.  \label{coltww76a}
\end{equation}%
We expand $\omega _{2}$ according to the various eigenvalues of $N$ like%
\begin{equation}
\omega _{2}=\sum\limits_{l>0}\omega _{2}^{\left( l\right) },
\label{coltww77}
\end{equation}%
where $N\omega _{2}^{\left( l\right) }=l\omega _{2}^{\left( l\right) }$,
such that
\begin{equation}
N\omega _{2}=\sum\limits_{l>0}l\omega _{2}^{\left( l\right) }.
\label{coltww78}
\end{equation}%
Comparing (\ref{coltww76a}) with (\ref{coltww78}), we reach the conclusion
that the decomposition (\ref{coltww77}) induces a similar decomposition with
respect to $W_{A}^{\lambda \mu \nu \gamma |\kappa \sigma }$ and $U_{a}^{\mu
\nu \gamma |\kappa \beta \sigma }$
\begin{equation}
W_{A}^{\lambda \mu \nu \gamma |\kappa \sigma }=\sum\limits_{l>0}W_{A\left(
l-1\right) }^{\lambda \mu \nu \gamma |\kappa \sigma },\quad U_{a}^{\mu \nu
\gamma |\kappa \beta \sigma }=\sum\limits_{l>0}U_{a\left( l-1\right) }^{\mu
\nu \gamma |\kappa \beta \sigma }.  \label{coltww79}
\end{equation}%
Substituting (\ref{coltww79}) into (\ref{coltww76a}) and comparing the
resulting expression with (\ref{coltww78}), we obtain that
\begin{equation}
\omega _{2}^{\left( l\right) }=\frac{1}{8l}K_{\lambda \mu \nu \gamma |\kappa
\sigma }^{A}W_{A\left( l-1\right) }^{\lambda \mu \nu \gamma |\kappa \sigma }+%
\frac{1}{9l}F_{\mu \nu \gamma |\kappa \beta \sigma }^{a}U_{a\left(
l-1\right) }^{\mu \nu \gamma |\kappa \beta \sigma }+\partial _{\mu }\bar{v}%
_{(l)}^{\mu }.  \label{coltvprform}
\end{equation}%
Introducing (\ref{coltvprform}) in (\ref{coltww77}), we arrive at
\begin{equation}
\omega _{2}=K_{\lambda \mu \nu \gamma |\kappa \sigma }^{A}\bar{W}%
_{A}^{\lambda \mu \nu \gamma |\kappa \sigma }+F_{\mu \nu \gamma |\kappa
\beta \sigma }^{a}\bar{U}_{a}^{\mu \nu \gamma |\kappa \beta \sigma
}+\partial _{\mu }\bar{v}^{\mu },  \label{coltww81}
\end{equation}%
where
\begin{equation}
\bar{W}_{A}^{\lambda \mu \nu \gamma |\kappa \sigma }=\sum\limits_{l>0}\frac{1%
}{8l}W_{A\left( l-1\right) }^{\lambda \mu \nu \gamma |\kappa \sigma },\quad
\bar{U}_{a}^{\mu \nu \gamma |\kappa \beta \sigma }=\sum\limits_{l>0}\frac{1}{%
9l}U_{a\left( l-1\right) }^{\mu \nu \gamma |\kappa \beta \sigma }.
\label{coltww82}
\end{equation}%
Applying $\gamma $ on (\ref{coltww81}), we infer that a necessary condition
for the existence of solutions to the equation $\gamma \omega _{2}=\partial
_{\mu }j_{2}^{\mu }$ is that the functions $\bar{W}_{A}^{\lambda \mu \nu
\gamma |\kappa \sigma }$ and $\bar{U}_{a}^{\mu \nu \gamma |\kappa \beta
\sigma }$ entering (\ref{coltww81}) must satisfy the equations
\begin{eqnarray}
\partial _{\rho }\left( F_{\mu \nu \gamma |\kappa \beta \sigma }^{b}\frac{%
\partial \bar{U}_{b}^{\mu \nu \gamma |\kappa \beta \sigma }}{\partial
r_{\rho \delta |\xi \chi }^{a}}+K_{\lambda \mu \nu \gamma |\kappa \sigma
}^{B}\frac{\partial \bar{W}_{B}^{\lambda \mu \nu \gamma |\kappa \sigma }}{%
\partial r_{\rho \delta |\xi \chi }^{a}}\right) &=&0,  \label{colrt80} \\
\partial _{\chi }\left( F_{\mu \nu \gamma |\kappa \beta \sigma }^{b}\frac{%
\partial \bar{U}_{b}^{\mu \nu \gamma |\kappa \beta \sigma }}{\partial
t_{\rho \delta \xi |\chi }^{A}}+K_{\lambda \mu \nu \gamma |\kappa \sigma
}^{B}\frac{\partial \bar{W}_{B}^{\lambda \mu \nu \gamma |\kappa \sigma }}{%
\partial t_{\rho \delta \xi |\chi }^{A}}\right) &=&0,  \label{colrt81} \\
\partial _{\rho }\left( F_{\mu \nu \gamma |\kappa \beta \sigma }^{b}\frac{%
\partial \bar{U}_{b}^{\mu \nu \gamma |\kappa \beta \sigma }}{\partial
t_{\rho \delta \xi |\chi }^{A}}+K_{\lambda \mu \nu \gamma |\kappa \sigma
}^{B}\frac{\partial \bar{W}_{B}^{\lambda \mu \nu \gamma |\kappa \sigma }}{%
\partial t_{\rho \delta \xi |\chi }^{A}}\right) &=&0.  \label{colrt82}
\end{eqnarray}%
The general solution to equations (\ref{colrt80})--(\ref{colrt82}) reads as%
\begin{eqnarray}
F_{\mu \nu \gamma |\kappa \beta \sigma }^{b}\frac{\partial \bar{U}_{b}^{\mu
\nu \gamma |\kappa \beta \sigma }}{\partial r_{\rho \delta |\xi \chi }^{a}}%
+K_{\lambda \mu \nu \gamma |\kappa \sigma }^{B}\frac{\partial \bar{W}%
_{B}^{\lambda \mu \nu \gamma |\kappa \sigma }}{\partial r_{\rho \delta |\xi
\chi }^{a}} &=&\partial _{\tau }\partial _{\theta }E_{a}^{\rho \delta \tau
|\xi \chi \theta },  \label{colrt83} \\
F_{\mu \nu \gamma |\kappa \beta \sigma }^{b}\frac{\partial \bar{U}_{b}^{\mu
\nu \gamma |\kappa \beta \sigma }}{\partial t_{\rho \delta \xi |\chi }^{A}}%
+K_{\lambda \mu \nu \gamma |\kappa \sigma }^{B}\frac{\partial \bar{W}%
_{B}^{\lambda \mu \nu \gamma |\kappa \sigma }}{\partial t_{\rho \delta \xi
|\chi }^{A}} &=&\partial _{\tau }\partial _{\theta }H_{A}^{\rho \delta \xi
\tau |\chi \theta },  \label{colrt84}
\end{eqnarray}%
where the functions $E_{a}^{\rho \delta \tau |\xi \chi \theta }$ and $%
H_{A}^{\rho \delta \xi \tau |\chi \theta }$ are derivative-free and exhibit
the mixed symmetries $(3,3)$ and $(4,2)$ respectively. By direct
computations we deduce%
\begin{eqnarray}
\partial _{\tau }\partial _{\theta }E_{a}^{\rho \delta \tau |\xi \chi \theta
} &=&\frac{\partial ^{2}E_{a}^{\rho \delta \tau |\xi \chi \theta }}{\partial
r_{\rho ^{\prime }\delta ^{\prime }|\xi ^{\prime }\chi ^{\prime
}}^{b}\partial r_{\rho ^{\prime \prime }\delta ^{\prime \prime }|\xi
^{\prime \prime }\chi ^{\prime \prime }}^{c}}\left( \partial _{\theta
}r_{\rho ^{\prime }\delta ^{\prime }|\xi ^{\prime }\chi ^{\prime
}}^{b}\right) \left( \partial _{\tau }r_{\rho ^{\prime \prime }\delta
^{\prime \prime }|\xi ^{\prime \prime }\chi ^{\prime \prime }}^{c}\right)
\notag \\
&&+\frac{\partial ^{2}E_{a}^{\rho \delta \tau |\xi \chi \theta }}{\partial
t_{\rho ^{\prime }\delta ^{\prime }\xi ^{\prime }|\chi ^{\prime
}}^{B}\partial t_{\rho ^{\prime \prime }\delta ^{\prime \prime }\xi ^{\prime
\prime }|\chi ^{\prime \prime }}^{C}}\left( \partial _{\theta }t_{\rho
^{\prime }\delta ^{\prime }\xi ^{\prime }|\chi ^{\prime }}^{B}\right) \left(
\partial _{\tau }t_{\rho ^{\prime \prime }\delta ^{\prime \prime }\xi
^{\prime \prime }|\chi ^{\prime \prime }}^{C}\right)  \notag \\
&&+\frac{\partial ^{2}\left( E_{a}^{\rho \delta \tau |\xi \chi \theta
}+E_{a}^{\rho \delta \theta |\xi \chi \tau }\right) }{\partial r_{\rho
^{\prime }\delta ^{\prime }|\xi ^{\prime }\chi ^{\prime }}^{b}\partial
t_{\rho ^{\prime \prime }\delta ^{\prime \prime }\xi ^{\prime \prime }|\chi
^{\prime \prime }}^{B}}\left( \partial _{\theta }r_{\rho ^{\prime }\delta
^{\prime }|\xi ^{\prime }\chi ^{\prime }}^{b}\right) \left( \partial _{\tau
}t_{\rho ^{\prime \prime }\delta ^{\prime \prime }\xi ^{\prime \prime }|\chi
^{\prime \prime }}^{B}\right)  \notag \\
&&+\frac{\partial E_{a}^{\rho \delta \tau |\xi \chi \theta }}{\partial
r_{\rho ^{\prime }\delta ^{\prime }|\xi ^{\prime }\chi ^{\prime }}^{b}}%
\partial _{\tau }\partial _{\theta }r_{\rho ^{\prime }\delta ^{\prime }|\xi
^{\prime }\chi ^{\prime }}^{b}+\frac{\partial E_{a}^{\rho \delta \tau |\xi
\chi \theta }}{\partial t_{\rho ^{\prime }\delta ^{\prime }\xi ^{\prime
}|\chi ^{\prime }}^{B}}\partial _{\tau }\partial _{\theta }t_{\rho ^{\prime
}\delta ^{\prime }\xi ^{\prime }|\chi ^{\prime }}^{B},  \label{colrt88}
\end{eqnarray}%
\begin{eqnarray}
\partial _{\tau }\partial _{\theta }H_{A}^{\rho \delta \xi \tau |\chi \theta
} &=&\frac{\partial ^{2}H_{A}^{\rho \delta \xi \tau |\chi \theta }}{\partial
r_{\rho ^{\prime }\delta ^{\prime }|\xi ^{\prime }\chi ^{\prime
}}^{b}\partial r_{\rho ^{\prime \prime }\delta ^{\prime \prime }|\xi
^{\prime \prime }\chi ^{\prime \prime }}^{c}}\left( \partial _{\theta
}r_{\rho ^{\prime }\delta ^{\prime }|\xi ^{\prime }\chi ^{\prime
}}^{b}\right) \left( \partial _{\tau }r_{\rho ^{\prime \prime }\delta
^{\prime \prime }|\xi ^{\prime \prime }\chi ^{\prime \prime }}^{c}\right)
\notag \\
&&+\frac{\partial ^{2}H_{A}^{\rho \delta \xi \tau |\chi \theta }}{\partial
t_{\rho ^{\prime }\delta ^{\prime }\xi ^{\prime }|\chi ^{\prime
}}^{B}\partial t_{\rho ^{\prime \prime }\delta ^{\prime \prime }\xi ^{\prime
\prime }|\chi ^{\prime \prime }}^{C}}\left( \partial _{\theta }t_{\rho
^{\prime }\delta ^{\prime }\xi ^{\prime }|\chi ^{\prime }}^{B}\right) \left(
\partial _{\tau }t_{\rho ^{\prime \prime }\delta ^{\prime \prime }\xi
^{\prime \prime }|\chi ^{\prime \prime }}^{C}\right)  \notag \\
&&+\frac{\partial ^{2}\left( H_{A}^{\rho \delta \xi \tau |\chi \theta
}+H_{A}^{\rho \delta \xi \theta |\chi \tau }\right) }{\partial r_{\rho
^{\prime }\delta ^{\prime }|\xi ^{\prime }\chi ^{\prime }}^{b}\partial
t_{\rho ^{\prime \prime }\delta ^{\prime \prime }\xi ^{\prime \prime }|\chi
^{\prime \prime }}^{B}}\left( \partial _{\theta }r_{\rho ^{\prime }\delta
^{\prime }|\xi ^{\prime }\chi ^{\prime }}^{b}\right) \left( \partial _{\tau
}t_{\rho ^{\prime \prime }\delta ^{\prime \prime }\xi ^{\prime \prime }|\chi
^{\prime \prime }}^{B}\right)  \notag \\
&&+\frac{\partial H_{A}^{\rho \delta \xi \tau |\chi \theta }}{\partial
r_{\rho ^{\prime }\delta ^{\prime }|\xi ^{\prime }\chi ^{\prime }}^{b}}%
\partial _{\tau }\partial _{\theta }r_{\rho ^{\prime }\delta ^{\prime }|\xi
^{\prime }\chi ^{\prime }}^{b}+\frac{\partial H_{A}^{\rho \delta \xi \tau
|\chi \theta }}{\partial t_{\rho ^{\prime }\delta ^{\prime }\xi ^{\prime
}|\chi ^{\prime }}^{B}}\partial _{\tau }\partial _{\theta }t_{\rho ^{\prime
}\delta ^{\prime }\xi ^{\prime }|\chi ^{\prime }}^{B}.  \label{colrt89}
\end{eqnarray}%
Substituting (\ref{colrt88})--(\ref{colrt89}) in (\ref{colrt83})--(\ref%
{colrt84}) and comparing the left-hand sides with the corresponding
right-hand sides of the resulting relations, we find the necessary equations%
\begin{eqnarray}
\frac{\partial ^{2}E_{a}^{\rho \delta \tau |\xi \chi \theta }}{\partial
r_{\rho ^{\prime }\delta ^{\prime }|\xi ^{\prime }\chi ^{\prime
}}^{b}\partial r_{\rho ^{\prime \prime }\delta ^{\prime \prime }|\xi
^{\prime \prime }\chi ^{\prime \prime }}^{c}} &=&0,\quad \frac{\partial
^{2}E_{a}^{\rho \delta \tau |\xi \chi \theta }}{\partial t_{\rho ^{\prime
}\delta ^{\prime }\xi ^{\prime }|\chi ^{\prime }}^{B}\partial t_{\rho
^{\prime \prime }\delta ^{\prime \prime }\xi ^{\prime \prime }|\chi ^{\prime
\prime }}^{C}}=0,  \label{colrt90} \\
\frac{\partial ^{2}H_{A}^{\rho \delta \xi \tau |\chi \theta }}{\partial
r_{\rho ^{\prime }\delta ^{\prime }|\xi ^{\prime }\chi ^{\prime
}}^{b}\partial r_{\rho ^{\prime \prime }\delta ^{\prime \prime }|\xi
^{\prime \prime }\chi ^{\prime \prime }}^{c}} &=&0,\quad \frac{\partial
^{2}H_{A}^{\rho \delta \xi \tau |\chi \theta }}{\partial t_{\rho ^{\prime
}\delta ^{\prime }\xi ^{\prime }|\chi ^{\prime }}^{B}\partial t_{\rho
^{\prime \prime }\delta ^{\prime \prime }\xi ^{\prime \prime }|\chi ^{\prime
\prime }}^{C}}=0,  \label{colrt91} \\
\frac{\partial ^{2}\left( E_{a}^{\rho \delta \tau |\xi \chi \theta
}+E_{a}^{\rho \delta \theta |\xi \chi \tau }\right) }{\partial r_{\rho
^{\prime }\delta ^{\prime }|\xi ^{\prime }\chi ^{\prime }}^{b}\partial
t_{\rho ^{\prime \prime }\delta ^{\prime \prime }\xi ^{\prime \prime }|\chi
^{\prime \prime }}^{B}} &=&0,\quad \frac{\partial ^{2}\left( H_{A}^{\rho
\delta \xi \tau |\chi \theta }+H_{A}^{\rho \delta \xi \theta |\chi \tau
}\right) }{\partial r_{\rho ^{\prime }\delta ^{\prime }|\xi ^{\prime }\chi
^{\prime }}^{b}\partial t_{\rho ^{\prime \prime }\delta ^{\prime \prime }\xi
^{\prime \prime }|\chi ^{\prime \prime }}^{B}}=0.  \label{colrt92}
\end{eqnarray}%
The above relations allow us to write%
\begin{equation}
\frac{1}{2}\left( E_{a}^{\rho \delta \tau |\xi \chi \theta }+E_{a}^{\rho
\delta \theta |\xi \chi \tau }\right) =C_{ab}^{\rho \delta \tau |\xi \chi
\theta ;\rho ^{\prime }\delta ^{\prime }|\xi ^{\prime }\chi ^{\prime
}}r_{\rho ^{\prime }\delta ^{\prime }|\xi ^{\prime }\chi ^{\prime
}}^{b}+C_{aB}^{\rho \delta \theta |\xi \chi \tau ;\rho ^{\prime }\delta
^{\prime }\xi ^{\prime }|\chi ^{\prime }}t_{\rho ^{\prime }\delta ^{\prime
}\xi ^{\prime }|\chi ^{\prime }}^{B},  \label{colrt93}
\end{equation}%
\begin{equation}
\frac{1}{2}\left( H_{A}^{\rho \delta \xi \tau |\chi \theta }+H_{A}^{\rho
\delta \xi \theta |\chi \tau }\right) =\hat{C}_{Ab}^{\rho \delta \xi \tau
|\chi \theta ;\rho ^{\prime }\delta ^{\prime }|\xi ^{\prime }\chi ^{\prime
}}r_{\rho ^{\prime }\delta ^{\prime }|\xi ^{\prime }\chi ^{\prime }}^{b}+%
\hat{C}_{AB}^{\rho \delta \xi \tau |\chi \theta ;\rho ^{\prime }\delta
^{\prime }\xi ^{\prime }|\chi ^{\prime }}t_{\rho ^{\prime }\delta ^{\prime
}\xi ^{\prime }|\chi ^{\prime }}^{B},  \label{colrt94}
\end{equation}%
where the quantities denoted by $C$ or $\hat{C}$ are some non-derivative,
real tensors, with the expressions%
\begin{eqnarray}
C_{ab}^{\rho \delta \tau |\xi \chi \theta ;\rho ^{\prime }\delta ^{\prime
}|\xi ^{\prime }\chi ^{\prime }} &=&\tilde{C}_{ab}^{\rho \delta \tau |\xi
\chi \theta ;\rho ^{\prime }\delta ^{\prime }|\xi ^{\prime }\chi ^{\prime }}+%
\tilde{C}_{ab}^{\rho \delta \theta |\xi \chi \tau ;\rho ^{\prime }\delta
^{\prime }|\xi ^{\prime }\chi ^{\prime }},  \label{colrt95} \\
C_{aB}^{\rho \delta \theta |\xi \chi \tau ;\rho ^{\prime }\delta ^{\prime
}\xi ^{\prime }|\chi ^{\prime }} &=&\tilde{C}_{aB}^{\rho \delta \tau |\xi
\chi \theta ;\rho ^{\prime }\delta ^{\prime }\xi ^{\prime }|\chi ^{\prime }}+%
\tilde{C}_{aB}^{\rho \delta \theta |\xi \chi \tau ;\rho ^{\prime }\delta
^{\prime }\xi ^{\prime }|\chi ^{\prime }},  \label{colrt96} \\
\hat{C}_{Ab}^{\rho \delta \xi \tau |\chi \theta ;\rho ^{\prime }\delta
^{\prime }|\xi ^{\prime }\chi ^{\prime }} &=&\bar{C}_{Ab}^{\rho \delta \xi
\tau |\chi \theta ;\rho ^{\prime }\delta ^{\prime }|\xi ^{\prime }\chi
^{\prime }}+\bar{C}_{Ab}^{\rho \delta \xi \theta |\chi \tau ;\rho ^{\prime
}\delta ^{\prime }|\xi ^{\prime }\chi ^{\prime }},  \label{colrt97} \\
\hat{C}_{AB}^{\rho \delta \xi \tau |\chi \theta ;\rho ^{\prime }\delta
^{\prime }\xi ^{\prime }|\chi ^{\prime }} &=&\bar{C}_{AB}^{\rho \delta \xi
\tau |\chi \theta ;\rho ^{\prime }\delta ^{\prime }\xi ^{\prime }|\chi
^{\prime }}+\bar{C}_{AB}^{\rho \delta \xi \theta |\chi \tau ;\rho ^{\prime
}\delta ^{\prime }\xi ^{\prime }|\chi ^{\prime }}.  \label{colrt98}
\end{eqnarray}%
Wherever two sets of indices are connected by a semicolon, it is understood
that the corresponding tensor possesses independently the mixed symmetries
with respect to the former and respectively the latter set. On the other
hand, it is obvious that%
\begin{eqnarray}
\partial _{\tau }\partial _{\theta }E_{a}^{\rho \delta \tau |\xi \chi \theta
} &=&\frac{1}{2}\partial _{\tau }\partial _{\theta }\left( E_{a}^{\rho
\delta \tau |\xi \chi \theta }+E_{a}^{\rho \delta \theta |\xi \chi \tau
}\right) ,  \label{colrt99} \\
\partial _{\tau }\partial _{\theta }H_{A}^{\rho \delta \xi \tau |\chi \theta
} &=&\frac{1}{2}\partial _{\tau }\partial _{\theta }\left( H_{A}^{\rho
\delta \xi \tau |\chi \theta }+H_{A}^{\rho \delta \xi \theta |\chi \tau
}\right) ,  \label{colrt100}
\end{eqnarray}%
so equations (\ref{colrt83})--(\ref{colrt84}) become%
\begin{gather}
F_{\mu \nu \gamma |\kappa \beta \sigma }^{b}\frac{\partial \bar{U}_{b}^{\mu
\nu \gamma |\kappa \beta \sigma }}{\partial r_{\rho \delta |\xi \chi }^{a}}%
+K_{\lambda \mu \nu \gamma |\kappa \sigma }^{B}\frac{\partial \bar{W}%
_{B}^{\lambda \mu \nu \gamma |\kappa \sigma }}{\partial r_{\rho \delta |\xi
\chi }^{a}}=  \notag \\
=C_{ab}^{\rho \delta \tau |\xi \chi \theta ;\rho ^{\prime }\delta ^{\prime
}|\xi ^{\prime }\chi ^{\prime }}\partial _{\tau }\partial _{\theta }r_{\rho
^{\prime }\delta ^{\prime }|\xi ^{\prime }\chi ^{\prime }}^{b}+C_{aB}^{\rho
\delta \theta |\xi \chi \tau ;\rho ^{\prime }\delta ^{\prime }\xi ^{\prime
}|\chi ^{\prime }}\partial _{\tau }\partial _{\theta }t_{\rho ^{\prime
}\delta ^{\prime }\xi ^{\prime }|\chi ^{\prime }}^{B},  \label{colrt101} \\
F_{\mu \nu \gamma |\kappa \beta \sigma }^{b}\frac{\partial \bar{U}_{b}^{\mu
\nu \gamma |\kappa \beta \sigma }}{\partial t_{\rho \delta \xi |\chi }^{A}}%
+K_{\lambda \mu \nu \gamma |\kappa \sigma }^{B}\frac{\partial \bar{W}%
_{B}^{\lambda \mu \nu \gamma |\kappa \sigma }}{\partial t_{\rho \delta \xi
|\chi }^{A}}=  \notag \\
=\hat{C}_{Ab}^{\rho \delta \xi \tau |\chi \theta ;\rho ^{\prime }\delta
^{\prime }|\xi ^{\prime }\chi ^{\prime }}\partial _{\tau }\partial _{\theta
}r_{\rho ^{\prime }\delta ^{\prime }|\xi ^{\prime }\chi ^{\prime }}^{b}+\hat{%
C}_{AB}^{\rho \delta \xi \tau |\chi \theta ;\rho ^{\prime }\delta ^{\prime
}\xi ^{\prime }|\chi ^{\prime }}\partial _{\tau }\partial _{\theta }t_{\rho
^{\prime }\delta ^{\prime }\xi ^{\prime }|\chi ^{\prime }}^{B}.
\label{colrt102}
\end{gather}%
Taking the partial derivatives of equations (\ref{colrt101}) and (\ref%
{colrt102}) with respect to $\partial _{\tau }\partial _{\theta }r_{\rho
^{\prime }\delta ^{\prime }|\xi ^{\prime }\chi ^{\prime }}^{b}$ and $%
\partial _{\tau }\partial _{\theta }t_{\rho ^{\prime }\delta ^{\prime }\xi
^{\prime }|\chi ^{\prime }}^{B}$, we infer the relations%
\begin{eqnarray}
\frac{\partial \bar{U}_{b}^{\mu \nu \gamma |\kappa \beta \sigma }}{\partial
r_{\rho \delta |\xi \chi }^{a}} &=&k_{ba}^{\mu \nu \gamma |\kappa \beta
\sigma ;\rho \delta |\xi \chi },\quad \frac{\partial \bar{W}_{B}^{\lambda
\mu \nu \gamma |\kappa \sigma }}{\partial r_{\rho \delta |\xi \chi }^{a}}=%
\bar{k}_{Ba}^{\lambda \mu \nu \gamma |\kappa \sigma ;\rho \delta |\xi \chi },
\label{colrt103} \\
\frac{\partial \bar{U}_{b}^{\mu \nu \gamma |\kappa \beta \sigma }}{\partial
t_{\rho \delta \xi |\chi }^{A}} &=&\hat{k}_{bA}^{\mu \nu \gamma |\kappa
\beta \sigma ;\rho \delta \xi |\chi },\quad \frac{\partial \bar{W}%
_{B}^{\lambda \mu \nu \gamma |\kappa \sigma }}{\partial t_{\rho \delta \xi
|\chi }^{A}}=\tilde{k}_{BA}^{\lambda \mu \nu \gamma |\kappa \sigma ;\rho
\delta \xi |\chi },  \label{colrt104}
\end{eqnarray}%
where $k_{ab}^{\mu \nu \gamma |\kappa \beta \sigma ;\rho \delta |\xi \chi }$%
,\ $\bar{k}_{aB}^{\lambda \mu \nu \gamma |\kappa \sigma ;\rho \delta |\xi
\chi }$,\ $\hat{k}_{Ab}^{\mu \nu \gamma |\kappa \beta \sigma ;\rho \delta
\xi |\chi }$, and $\tilde{k}_{AB}^{\lambda \mu \nu \gamma |\kappa \sigma
;\rho \delta \xi |\chi }$ denote some non-derivative, constant tensors. By
means of relations (\ref{colrt103}) and (\ref{colrt104}) we obtain (up to
some irrelevant constants)%
\begin{eqnarray}
\bar{U}_{b}^{\mu \nu \gamma |\kappa \beta \sigma } &=&k_{ba}^{\mu \nu \gamma
|\kappa \beta \sigma ;\rho \delta |\xi \chi }r_{\rho \delta |\xi \chi }^{a}+%
\hat{k}_{bA}^{\mu \nu \gamma |\kappa \beta \sigma ;\rho \delta \xi |\chi
}t_{\rho \delta \xi |\chi }^{A},  \label{colrt105} \\
\bar{W}_{B}^{\lambda \mu \nu \gamma |\kappa \sigma } &=&\bar{k}%
_{Ba}^{\lambda \mu \nu \gamma |\kappa \sigma ;\rho \delta |\xi \chi }r_{\rho
\delta |\xi \chi }^{a}+\tilde{k}_{BA}^{\lambda \mu \nu \gamma |\kappa \sigma
;\rho \delta \xi |\chi }t_{\rho \delta \xi |\chi }^{A}.  \label{colrt106}
\end{eqnarray}%
From the expression of $\omega _{2}$ given by (\ref{coltww81}) we notice
that the terms $k_{ba}^{\mu \nu \gamma |\kappa \beta \sigma ;\rho \delta
|\xi \chi }r_{\rho \delta |\xi \chi }^{a}$ and $\tilde{k}_{BA}^{\lambda \mu
\nu \gamma |\kappa \sigma ;\rho \delta \xi |\chi }t_{\rho \delta \xi |\chi
}^{A}$ appearing in (\ref{colrt105}) and (\ref{colrt106}) bring no
contributions to cross-interactions. For this reason, we take%
\begin{equation}
k_{ba}^{\mu \nu \gamma |\kappa \beta \sigma ;\rho \delta |\xi \chi }=0,\quad
\tilde{k}_{BA}^{\lambda \mu \nu \gamma |\kappa \sigma ;\rho \delta \xi |\chi
}=0,  \label{colrt107}
\end{equation}%
such that (up to a total, irrelevant divergence) $\omega _{2}$ takes the form%
\begin{equation}
\omega _{2}=\bar{k}_{Aa}^{\lambda \mu \nu \gamma |\kappa \sigma ;\rho \delta
|\xi \chi }K_{\lambda \mu \nu \gamma |\kappa \sigma }^{A}r_{\rho \delta |\xi
\chi }^{a}+\hat{k}_{aA}^{\mu \nu \gamma |\kappa \beta \sigma ;\rho \delta
\xi |\chi }F_{\mu \nu \gamma |\kappa \beta \sigma }^{a}t_{\rho \delta \xi
|\chi }^{A}.  \label{colrt108}
\end{equation}%
The most general expression of $\bar{k}_{Aa}^{\lambda \mu \nu \gamma |\kappa
\sigma ;\rho \delta |\xi \chi }$ is represented by%
\begin{eqnarray}
\bar{k}_{Aa}^{\lambda \mu \nu \gamma |\kappa \sigma ;\rho \delta |\xi \chi }
&=&c_{Aa}\left[ \frac{1}{4}\varepsilon ^{\lambda \mu \nu \gamma \rho \delta
}\left( \sigma ^{\xi \kappa }\sigma ^{\chi \sigma }-\sigma ^{\xi \sigma
}\sigma ^{\chi \kappa }\right) \right.  \notag \\
&&+\frac{1}{4}\varepsilon ^{\lambda \mu \nu \gamma \xi \chi }\left( \sigma
^{\rho \kappa }\sigma ^{\delta \sigma }-\sigma ^{\rho \sigma }\sigma
^{\delta \kappa }\right)  \notag \\
&&\left. -\frac{1}{24}\varepsilon ^{\lambda \mu \nu \gamma \left[ \rho
\delta \right. }\delta _{\tau }^{\xi }\delta _{\theta }^{\left. \chi \right]
}\left( \sigma ^{\tau \kappa }\sigma ^{\theta \sigma }-\sigma ^{\tau \sigma
}\sigma ^{\theta \kappa }\right) \right] ,  \label{colrt109}
\end{eqnarray}%
which then yields
\begin{equation}
\bar{k}_{Aa}^{\lambda \mu \nu \gamma |\kappa \sigma ;\rho \delta |\xi \chi
}K_{\lambda \mu \nu \gamma |\kappa \sigma }^{A}r_{\rho \delta |\xi \chi
}^{a}=c_{Aa}\varepsilon ^{\lambda \mu \nu \gamma \rho \delta }r_{\rho \delta
|\xi \chi }^{a}K_{\lambda \mu \nu \gamma |}^{A\quad \ \xi \chi },
\label{colrt110}
\end{equation}%
with $c_{Aa}$ some real constants. On the other hand, there exist
non-trivial constant tensors of the type $\hat{k}_{aA}^{\mu \nu \gamma
|\kappa \beta \sigma ;\rho \delta \xi |\chi }$, but they all lead in the end
to%
\begin{equation}
\hat{k}_{aA}^{\mu \nu \gamma |\kappa \beta \sigma ;\rho \delta \xi |\chi
}F_{\mu \nu \gamma |\kappa \beta \sigma }^{a}t_{\rho \delta \xi |\chi
}^{A}\equiv 0  \label{colrt111}
\end{equation}%
due to the algebraic Bianchi I identities $F_{\left[ \mu \nu \gamma |\kappa %
\right] \beta \sigma }^{a}\equiv 0$. Such constants have an intricate and
non-illuminating form, and therefore we will skip them. Inserting (\ref%
{colrt110}) and (\ref{colrt111}) in (\ref{colrt108}), we deduce%
\begin{equation}
\omega _{2}=c_{Aa}\varepsilon ^{\lambda \mu \nu \gamma \rho \delta }r_{\rho
\delta |\xi \chi }^{a}K_{\lambda \mu \nu \gamma |}^{A\quad \ \xi \chi }.
\label{colrt112}
\end{equation}%
Acting with $\gamma $ on (\ref{colrt112}), it is easy to see that%
\begin{equation}
\gamma \omega _{2}=-2c_{Aa}\varepsilon ^{\lambda \mu \nu \gamma \rho \delta
}\left( \partial _{\left[ \lambda \right. }K_{\mu \nu \gamma ]|}^{A\quad \xi
}\right) \mathcal{C}_{\rho \delta |\xi }^{a}\neq \partial _{\mu }j_{2}^{\mu
},  \label{colrt113}
\end{equation}%
where $K_{\mu \nu \gamma |\tau }^{A}$ is the trace of the curvature tensor $%
K_{\mu \nu \gamma \kappa |\tau \beta }^{A}$, $K_{\mu \nu \gamma |\tau
}^{A}=\sigma ^{\kappa \beta }K_{\mu \nu \gamma \kappa |\tau \beta }^{A}$. It
is worthy to notice that $\gamma \omega _{2}\neq \partial _{\mu }j_{2}^{\mu
} $ follows from the differential Bianchi II identity $\partial _{\beta
}K_{\lambda \mu \nu \gamma |}^{A\quad \ \beta \xi }=\partial _{\left[
\lambda \right. }K_{\mu \nu \gamma ]|}^{A\quad \xi }$. Due to (\ref{colrt113}%
), we must take%
\begin{equation}
c_{Aa}=0,  \label{colrt114}
\end{equation}%
and hence%
\begin{equation}
\omega _{2}=0.  \label{colomega2}
\end{equation}%
Replacing (\ref{colomega0}), (\ref{colomega1}), and (\ref{colomega2}) in (%
\ref{coltww60}), we finally find (\ref{coltr102}).

\end{document}